\documentclass[pdflatex,sn-mathphys-num]{sn-jnl}

\usepackage{graphicx}%
\usepackage{multirow}%
\usepackage{amsmath,amssymb,amsfonts}%
\usepackage{amsthm}%
\usepackage{mathrsfs}%
\usepackage[title]{appendix}%
\usepackage{xcolor}%
\usepackage{textcomp}%
\usepackage{manyfoot}%
\usepackage{booktabs}%
\usepackage{multirow}%
\usepackage{tabularx}%
\usepackage{algorithm}%
\usepackage{algorithmicx}%
\usepackage{algpseudocode}%
\usepackage{listings}%
\usepackage{placeins}
\usepackage{makecell}

\theoremstyle{thmstyleone}%
\theoremstyle{thmstyletwo}%

\theoremstyle{thmstylethree}%

\begin{document}

\title[Neuromorphic Computing in Radio Observatories]{The Potential Impact of Neuromorphic Computing on Radio Telescope Observatories}


\author*[1,2]{\fnm{Nicholas J.} \sur{Pritchard}}\email{nicholas.pritchard@research.uwa.edu.au}

\author[1]{\fnm{Richard} \sur{Dodson}}\email{richard.dodson@icrar.org}
\equalcont{These authors contributed equally to this work.}

\author[1]{\fnm{Andreas} \sur{Wicenec}}\email{andreas.wicenec@uwa.edu.au}
\equalcont{These authors contributed equally to this work.}

\affil*[1]{\orgdiv{International Centre for Radio Astronomy Research}, \orgname{University of Western Australia}, \orgaddress{\street{7 Fairway}, \city{Perth}, \postcode{6009}, \state{WA}, \country{Australia}}}

\affil[2]{\orgdiv{School of Physics, Mathematics and Computing}, \orgname{University of Western Australia}, \orgaddress{\street{35 Stirling Highway}, \city{Perth}, \postcode{6009}, \state{WA}, \country{Australia}}}


\abstract{Radio astronomy relies on bespoke, experimental and innovative computing solutions. 
This will continue as next-generation telescopes such as the Square Kilometre Array (SKA) and next-generation Very Large Array (ngVLA) take shape.
Under increasingly demanding power consumption, and increasingly challenging radio environments, science goals may become intractable with conventional von Neumann computing due to related power requirements.
Neuromorphic computing offers a compelling alternative, and combined with a desire for data-driven methods, Spiking Neural Networks (SNNs) are a promising real-time power-efficient alternative.
Radio Frequency Interference (RFI) detection is an attractive use-case for SNNs where recent exploration holds promise.
This work presents a comprehensive analysis of the potential impact of deploying varying neuromorphic approaches across key stages in radio astronomy processing pipelines for several existing and near-term instruments.
Our analysis paves a realistic path from near-term FPGA deployment of SNNs in existing instruments, allowing the addition of advanced data-driven RFI detection for no capital cost, to neuromorphic ASICs for future instruments, finding that commercially available solutions could reduce the power budget for key processing elements by up to three orders of magnitude, transforming the operational budget of the observatory. 
High-data-rate spectrographic processing could be a well-suited target for the neuromorphic computing industry, as we cast radio telescopes as the world's largest in-sensor compute challenge.}

\keywords{radio astronomy, spiking neural networks, neuromorphic computing, non-Von Neumann computing}



\maketitle

\section{Introduction}\label{sec:introduction}
Astronomy is among the oldest, most storied fields in science, and has always been defined by its instruments.
Radio astronomy, by extension, has always been limited by the computing capacity of any facilities associated with an observatory.
Fundamentally, radio telescopes are faced with a complex, multi-dimensional data reduction problem.
Radio emission (from anthropogenic and cosmic sources) causes small electric fields in receivers, which are digitized; these raw voltages are correlated into spectrographic data (visibilities).
The first processing step involves detecting and eliminating anthropogenic radio emissions, known as Radio Frequency Interference (RFI), which would otherwise overwhelm subsequent processing steps.
These visibilities are subsequently averaged and carefully accumulated by iteratively removing bright foreground artifacts while considering the Earth's curvature, ionospheric interference, instrument noise, and a myriad of other physical phenomena \cite{thompson_interferometry_2017}.
To complicate matters, simultaneously detection fast-changing `transient' events like Pulsars is a core scientific goal for modern instruments, as these baffling objects represent some of the most extreme examples of physics ever observed \cite{hurley-walker_radio_2022}.
We outline this process to make clear that, at its core, radio astronomy is an in-sensor compute challenge, which, for convenience, has been treated with a batch processing approach.
It is in many ways better to think of radio telescopes as microphones, rather than cameras, and it is through this lens that we see radio astronomy as a field primed for neuromorphic computing.

\subsection{The Data Deluge of Radio Astronomy}
Near-term instruments are faced with an order of magnitude (at least) leap in sensitivity, survey speed and, correspondingly, data-processing challenges \cite{vermij_challenges_2015}. 
The initial data rates of the SKA-Low, SKA-Mid, and the ngVLA, for example, place these instruments firmly in high-performance computing (HPC) territory but known processing goals, providing the motivation to create highly optimized approaches to reach them \cite{fiorin_exploring_2016}.
In light of this incoming scale, the field is moving to a High Throughput Computing (HTC) or Many-Task Computing (MTC) where processing tasks are increasingly granular, increasingly parallelized and performed on increasingly heterogeneous resources \cite{raicu_many-task_2008, wu_daliuge:_2017}. 
No instrument can process anywhere near its maximum observational capacity in real-time \cite{vermij_challenges_2015}.
The bottleneck, as is the case in many other domains, is data movement at a micro and macro level.
The more efficient an instrument's computing platform is, the more science can be accomplished within a given operational budget.
Confounding this challenge further is the conflicting needs of scientific observations.
Transient events, such as Fast Radio Bursts (FRBs) and pulsars, are among the most scientifically interesting objects to observe, but require real-time detection.
Separate processing pipelines, explicitly designed for rapid coarse imaging and detecting these dynamic phenomena (sometimes with a frequency resolution of milliseconds), are an element of science enabled by high-performance computing.
Conversely, survey science, which involves scheduled observation of sections of the sky at particular times and conditions, is often more flexible, but only to certain limits. 
Epoch of Reionization (EoR) observations are the hallmark science case for large interferometric radio telescopes, such as the SKA, revealing a period of history that has never been seen before \cite{schilizzi_evolution_2024}.
These observations operate at the highest limit of sensitivity possible and require the largest quantity of data to provide sufficiently accurate images.
Very-long Baseline Interferometry (VLBI) utilizes multiple telescopes around the world in concert to image very distant objects, such as quasars and black holes, which are otherwise impossible to image with conventional techniques \cite{collaboration_first_2019}.
VLBI is also `turned around' to provide astronometry and geodesy measurements of the Earth \cite{sovers_astrometry_1998}.
Such observations require accurate modeling of ionospheric effects and measurements over multiple time periods at accurately determined intervals.
Each observation type has different processing requirements and cadences, creating a complex scheduling challenge in which speed and efficiency directly correlate with greater science throughput.

\subsection{Neuromorphic Computing is a Paradigm Shift}
Designing brain-inspired computing architectures has long promised significant advantages in power consumption and data transfer \cite{schuman_opportunities_2022}.
Neuromorphic computing also has a long history of struggling to find suitable use-cases beyond neuroscience.
However, this is starting to change with the advent of commercially available digital or mixed-signal accelerators \cite{muir_road_2025}.
Application domains where information is sparse, and real-time operation under energy constraints is beneficial hold particular promise for SNNs in particular \cite{eshraghian_training_2022}.
We argue that neuromorphic computing in the form of brain-inspired algorithms, FPGA hardware implementations of Hebbian style \cite{heidarpur_cordic-snn_2019} and backpropagation-style learning rules \cite{pham_review_2021,karamimanesh_spiking_2025}, custom digital ASICs \cite{frenkel_bottom-up_2023} could all find a place in radio astronomy, at various stages of the processing pipeline, using different neuromorphic hardware approaches.

Being an observational science built around processing time-varying data, ideally in real-time, makes this domain a natural fit for neuromorphic systems.
Moreover, while the associated data processing scale is massive, the data is relatively sparse; intuitively, there is a lot of nothing in space.
Systems that scale power usage around information density are key, and this is where neuromorphic computing may provide a unique advantage.
The unique combination of vast spectrographic data, real-time receiver voltage processing, the 24/7 potential for observation, and a history of methodological experimentation at the limit of technology makes radio astronomy an ideal field for addressing many engineering challenges in neuromorphic systems.

The initial receiving, correlation, RFI flagging, transient detection, and pulsar detection can all, in principle, be implemented as real-time processing steps.
Additionally, the volume of data and increased sensitivity of incoming instruments encourages data-driven, ML methods to tackle these tasks.
The operational cost of implementing contemporary deep-learning approaches often precludes their practical deployment \cite{dutoit_comparison_2024}.
Therefore, exploring Spiking Neural Networks (SNNs) holds unique promise for enabling data-driven, real-time signal processing tasks.
Imaging, however, is a different challenge where contemporary methods require iterative processing along orthogonal data dimensions (frequency and time) \cite{thompson_interferometry_2017}; a nightmare for data placement that is firmly limited by the von Neumann bottleneck \cite{corda_near_2020}.

\subsection{Related Work}
Initial work into applying neuromorphic computing to RFI detection has specifically shown promise.
Traditional algorithms are functional but require expert tuning, while Artificial Neural Network (ANN) data-driven methods are often prohibitively expensive to run operationally \cite{dutoit_comparison_2024}.
Radio astronomy has been known to be possible use-case for SNNs for some time \cite{kasabov_evolving_2016} with a very preliminary work testing pulsar detection in purely synthetic data appearing in the appendix of a thesis \cite{scott_evolving_2015}.
Moreover, SNNs have found use in other high-data rate filtering challenges, such as those found in High-Energy Physics \cite{miniskar_neuro-spark_2024}, where efficient implementation in FPGA logic is especially promising, and, in the filtering of interference in satellite communications \cite{eappen_neuromorphic_2025}.
A recent series of works aimed to scope out the feasibility of applying neuromorphic computing (SNNs specifically) to RFI detection, effectively testing out the time-varying data to time-varying compute platform prior.
RFI detection is an ideal candidate task for several reasons.
Firstly, all telescopes must flag RFI, unless they are located off-Earth.
Second, the challenge of mitigating RFI has grown significantly in recent years, mainly due to the presence of low-Earth orbit satellites.
Third, machine-learning ANN-based techniques exist, but are considered too expensive to deploy operationally, often retaining the same image-centric approach of conventional algorithms.
Finally, multiple flagging methods can be used in conjunction with each other, permitting some degree of flexibility and a tolerance for experimentation, compared to other aspects of radio astronomy processing, such as imaging.
Initial work applied ANN-to-SNN conversion, keeping the same image-centric approach as traditional methods but lowering a linearly growing memory bound to a constant at inference time \cite{pritchard_rfi_2024}.
The same team then moved to training SNNs with Backpropagation Through Time (BPTT) for the same task, finding comparable performance and exploring ideal spike-encoding techniques \cite{pritchard_supervised_2024}, later expanding that work to include real LOFAR-based results \cite{pritchard_spiking_2024}.
More exotic ideas have been trialed, including liquid state machines \cite{pritchard_advancing_2025}.
Including polarization information improves detection performance \cite{pritchard_polarisation-inclusive_2025}, and deploying these methods on SynSense Xylo hardware yields RFI detection in real-time at less than 50mW per baseline \cite{pritchard_neuromorphic_2025}.
This work has scoped, de-risked and explored the applicability of SNNs and neuromorphic computing in this processing step, and it is here we discuss the practical possibility and impact deploying such solutions could have.

A system-level, forward-looking analysis of the impact across the entire processing pipeline for next-generation observatories is missing.
In this work, we sketch out a path to adopting neuromorphic technologies into radio astronomy, starting with the most straightforward applications and least neuromorphic methods, then building into increasingly complex and neuromorphic applications, outlining the potential for energy savings along the way, with concrete modeling on existing and imminent telescopes where available.
\section{Results}\label{sec:results}
We now present modeling of hypothetical deployments of neuromorphic computing into several radio telescopes across RFI detection, transient searching and imaging.
Details about the telescopes under consideration are available in Section \ref{subsec:processing_models} as is a summary of their existing data processing resources in Section \ref{subsubsec:existing_hardware}. 
However, we briefly summarize them here:
The \textbf{Murchison Widefield Array (MWA)} is a low-frequency radio interferometer telescope located in remote Western Australia and is a precursor to the SKA-Low instrument comprised of 256 phased array `tiles' of 16 dipole antennas each \cite{lonsdale_murchison_2009}.
\textbf{Australian Square Kilometre Array Pathfinder (ASKAP)} is a dish-based radio telescope also located in remote Western Australia comprised of 36 12-meter parabolic antennas \cite{hotan_australian_2021}.
\textbf{The LOw Frequency ARray (LOFAR)} is another SKA-Low pathfinder based in the Netherlands with instrument components distributed across Europe \cite{van_haarlem_lofar_2013}.
The \textbf{SKA} is a global collaboration to build the world's largest radio telescope. The observatory is headquartered in Manchester, UK which operates two instruments in Australia and South Africa across two different frequency ranges, `Low' and `Mid' \cite{schilizzi_square_2024}.
The \textbf{SKA-Low} instrument is the low-frequency half of the SKA and is based in remote Western Australia.
The \textbf{SKA-Mid} instrument is the mid-frequency half of the SKA and is based in the Karoo in South Africa.
Finally, the \textbf{next-generation Very Large Array (ngVLA)} is an under-design replacement to the Very Large Array instrument. It will comprise of 282 dishes \cite{selina_ngvla_nodate}.
\subsection{RFI Flagging}
RFI flagging is an appealing first task to approach with neuromorphic methods as flagging can occur at many stages in a telescope's signal chain and even multiple times.
In an ever-increasingly noisy radio sky, there is strong motivation to use data-driven methods for this task at several stages in the processing pipeline.
Neuromorphic computing may offer an extremely frugal way to enable this.  
Table \ref{tab:dep:rfi} contains the expected benefits of deploying neuromorphic approaches in short, medium and long-term scenarios across multiple instruments.
Across several instruments, neuromorphic computing has the potential to provide RFI detection at one to three orders of magnitude less energy consumption; making data-driven, real-time RFI detection a feasible possibility.
The methods section outlines in significantly more detail our reasoning behind selecting quantities and choice of neuromorphic hardware for each instrument.
Our rationale for selecting an appropriate neuromorphic platform primarily focuses on lining up platforms with sufficient data transfer capabilities to handle the number of frequency channels provided by a particular instrument for that particular processing task.
In the case of correlator and post-correlator activities, we explore SpiNNaker 2 boards as they hold standard ARM processor cores in addition to SNN-acceleration, making these more numerical stages more plausible and feasible \cite{theilman_solving_2025}. 
\begin{table}[!htbp]
\centering\tiny
\caption{Hypothetical neuromorphic deployment scenarios for RFI detection in contemporary radio telescopes.}
\label{tab:dep:rfi}

\begin{tabularx}{\linewidth}{@{}cccccccc@{}}

\toprule

Instrument                & Location                     & \makecell{Current\\Hardware}              & \begin{tabular}[c]{@{}c@{}}Proposed\\ Hardware\end{tabular} & Quantity & \begin{tabular}[c]{@{}c@{}}Current\\ Power (W)\end{tabular} & \begin{tabular}[c]{@{}c@{}}Proposed\\ Power (W)\\(Relative \%)\end{tabular} & Timeline \\ \midrule

\multirow{6}{*}{MWA}     & \multirow{3}{*}{Receiving}   & \multirow{3}{*}{FPGA} & -                                                           & -        & \multirow{3}{*}{1024}                                       & -                                                                                                                        & Near     \\

                         &                              &                       & \makecell{SynSense\\Xylo 2}                                             & 2048     &                                                             & \makecell{1.23\\(0.12)}                                                                                                                 & Medium   \\

                         & Correlation                  & GPU                   & \makecell{Intel Loihi 2\\Oheo Gulch}                                    & 24       & 18720                                                       & \makecell{24\\(0.13)}                                                                                                                   & Far      \\

                         & Post-Correlator              & CPU/GPU               & SpiNNaker 2                                           & 304      & 152000                                                      & \makecell{7296\\(4.80)}                                                                                                                 & Far      \\ \midrule

\multirow{10}{*}{ASKAP}   & \multirow{3}{*}{Receiving}   & \multirow{3}{*}{FPGA} & -                                                           & -        & \multirow{3}{*}{3456}                                       & -                                                                                                                    & Near     \\

                         &                              &                       & \makecell{SynSense\\Xylo 2}                                             & 1728     &                                                             & \makecell{1.04\\(0.03)}                                                                                                              & Medium   \\

                         & \multirow{3}{*}{Correlation} & \multirow{3}{*}{FPGA} & -                                                           & -        & \multirow{3}{*}{49896}                                      & -                                                                                                                       & Near     \\

                         &                              &                       & \makecell{Intel Loihi 2\\Oheo Gulch}                                    & 252      &                                                             & \makecell{252\\(0.51)}                                                                                                                      & Medium   \\

                         & \makecell{Transient\\Detection}          & FPGA                  & -                                                           & -        & 4500                                                        & -                                                                                                                    & Near     \\

                         & Post-Correlator              & CPU/GPU               & SpiNNaker 2                                           & 304      & 152000                                                      & \makecell{7296\\(4.80)}                                                                                                            & Far      \\ \midrule

\multirow{5}{*}{LOFAR}   & \multirow{3}{*}{Receiving}   & \multirow{3}{*}{FPGA} & -                                                           & -        & \multirow{3}{*}{49920}                                      & -                                                                                                                    & Near     \\

                         &                              &                       & \makecell{Intel Loihi 2\\Oheo Gulch}                                    & 1664     &                                                             & \makecell{1664\\(3.33)}                                                                                                                & Medium   \\

                         & Correlation                  & GPU                   & SpiNNaker 2                                           & 26       & 10140                                                       & \makecell{624\\(6.15)}                                                                                                               & Far      \\

                         & Post-Correlator              & GPU                   & SpiNNaker 2                                           & 54       & 7280                                                        & \makecell{1296\\(17.80)}                                                                                                                  & Far      \\ \midrule

\multirow{3}{*}{SKA-LOW} & \multirow{3}{*}{Receiving}   & \multirow{3}{*}{FPGA} & -                                                           & -        & \multirow{3}{*}{49152}                                      & -                                                                                                                   & Near     \\

                         &                              &                       & \makecell{SynSense\\Xylo 2}                                             & 32768    &                                                             & \makecell{19.7\\(0.04)}                                                                                                                & Medium   \\ \midrule

\multirow{6}{*}{SKA-MID} & \multirow{3}{*}{Receiving}   & \multirow{3}{*}{FPGA} & -                                                           & -        & \multirow{3}{*}{14775}                                      & -                                                                                                                     & Near     \\

                         &                              &                       & \makecell{Intel Loihi 2\\Kapoho point}                                  & 197      &                                                             & \makecell{1576\\(10.67)}                                                                                                                  & Medium   \\

                         & \multirow{3}{*}{Correlation} & \multirow{3}{*}{FPGA} & -                                                           & -        & \multirow{3}{*}{13500}                                      & -                                                                                                                      & Near     \\

                         &                              &                       & \makecell{Intel Loihi 2\\Kapoho point}                                  & 180      &                                                             & \makecell{1440\\(10.67)}                                                                                                                  & Medium   \\ \midrule

\multirow{6}{*}{ngVLA}   & \multirow{3}{*}{Receiving}   & \multirow{3}{*}{FPGA} & -                                                           & -        & \multirow{3}{*}{19725}                                      & -                                                                                                                      & Near     \\

                         &                              &                       & \makecell{Intel Loihi 2\\Kapoho point}                                  & 263      &                                                             & \makecell{2104\\(10.67)}                                                                                                                & Medium   \\

                         & \multirow{3}{*}{Correlation} & \multirow{3}{*}{FPGA} & -                                                           & -        & \multirow{3}{*}{18000}                                      & -                                                                                                                       & Near     \\

                         &                              &                       & \makecell{Intel Loihi 2\\Kapoho point}                                  & 240      &                                                             & \makecell{1920\\(10.67)}                                                                                                                 & Medium   \\ \bottomrule

\end{tabularx}
\begin{tablenotes}
    \item Deployment scenarios are speculative by nature, but provide ample motivation to pursue further investigation into neuromorphic techniques in radio astronomy. Please refer to Section \ref{subsec:processing_models} for information about each telescope, Table \ref{tab:impact:meth:traditional} for information about each telescope's computing facilities, and Section \ref{subsec:neuro_hw} for information about the neuromorphic hardware options.
    \item Existing work on RFI-detection with SNNs has shown promise with both first and second-order Leaky Integrate and Fire (LiF) neurons \cite{pritchard_spiking_2024}, admitting a wide-range of hardware implementations to this particular problem.
\end{tablenotes}
\end{table}

Post-correlation RFI detection, which is the typical stage at which contemporary RFI detection methods are deployed has already been demonstrated as a first application for existing ASIC-based neuromorphic chipsets. 
Supplementary Table 1 presents several SNN-based approaches applied to this very problem for a representative but synthetic dataset and Supplementary Table 2 contains results for SNN baselines on a real LOFAR-derived dataset.
For instruments like ASKAP, SKA-LOW, SKA-MID, and ngVLA, which rely heavily on FPGA-processing, an SNN based approach to real-time RFI detection offers advantages primarily in model size while maintaining competitive performance.
Competitive results with first and second order LiF neurons under ANN2SNN \cite{pritchard_rfi_2024}, backpropagation through time \cite{pritchard_spiking_2024, pritchard_polarisation-inclusive_2025, pritchard_neuromorphic_2025}, and liquid state machine \cite{pritchard_advancing_2025} training approaches shows promise for this particular domain.

RFI detection is undeniably the most promising first potential use case of SNNs and neuromorphic computing in radio astronomy, having already shown promise in recent years.

These results are, of course, speculative; our intention is to show that contemporary neuromorphic hardware has the precedent and scale required to support this niche but data-intensive use case.
We see that the most straightforward approach to deploying neuromorphic approaches lies in utilizing the headroom available in existing FPGA resources across several instruments.
This is sensible for two main reasons, the capital cost is zero and the existing hardware is already capable of handling the required data-rates. 
One would hope that an initial investigation would motivate the deployment of available neuromorphic hardware at a capital refresh point, which happens every several years for sizable instruments. 
More fanciful applications are longer-term in nature and generally lie in later stages of the processing chain.

\subsection{Transient Detection}
Transients represent some of the least well-understood, and therefore most interesting, observable phenomena, making them a high priority science case for most instruments under operation and construction.
It is possible to search for transients in archived observations \cite{swinbank_lofar_2015, sokolowski_southern-hemisphere_2021}, however real-time transient detection is of particular interest for several modern instruments.
Table \ref{tab:dep:transient} contains a summary of the latency requirements for a hypothetical neuromorphic system for real-time transient detection across several instruments.

In this particular use-case the focus is on latency, and it is here where a time-varying SNN or neuromorphic approach to data processing has long been expected as advantageous \cite{scott_evolving_2015}.
Moreover, imminent instruments are expected to achieve sensitivity in excess of all prior observation, and therefore, adaptive systems capable of detection previously unobserved transients are advantageous too, and this is another arena where neuromorphic techniques will prove useful.
While not explored concretely as of yet, there is a strong expectation that SNN or neuromorphic techniques could hold significant impact in this high-priority science case.

\begin{table}[!htbp]
\centering
\caption{Neuromorphic deployment scenarios for transient detection.}
\label{tab:dep:transient}
\begin{tabularx}{\linewidth}{@{}ccccccccc@{}}
\toprule
Intrument & Channels & Latency & \begin{tabular}[c]{@{}c@{}}Proposed\\ Hardware\end{tabular}  & \begin{tabular}[c]{@{}c@{}}Current\\ Power (W)\end{tabular} & \begin{tabular}[c]{@{}c@{}}Proposed\\ Power (W)\end{tabular} & \begin{tabular}[c]{@{}c@{}}Power\\ Proportion (\%)\end{tabular} \\ \midrule
ASKAP          & 256        & 1.728ms & Loihi 2 Alia Point                                            & 4500                                                        & 128                                                          & 2.84                                                      \\
SKA-LOW        & 500        & 100$\mu$s   & Loihi 2 Hala Point                                            & 60000                                                       & 1152                                                         & 1.92                                                            \\ \bottomrule
\end{tabularx}
\end{table}
The `CRACO' system in ASKAP and its sequel in the SKA-LOW are of particular interest.
The ASKAP pipeline uses 20 Xilinx Alveo U280 FPGAs, and needs to handle 23 Tpixels per second at a latency of 1.728ms, effectively processing images of $256 \times 256$ size \cite{wang_craft_2025}.
A large asynchronous neuromorphic system may be able to handle such a task, so we consider an Intel Loihi 2 Alia Point system, consuming 128W which may be capable of handling the requisite 100Gbps incoming link and the minimum 200ns end-to-end response time may suffice for realtime operation.

The SKA-Low uses a combined correlator, beamformer and transient detection machine comprised of 400 Xilinx Alveo U55C at a combined 60kW of energy consumption \cite{hampson_square_2022}.
An Intel Loihi 2 Hala Point with 1.1kW of power usage may be up to the stringent latency and data-volume requirements.
The current correlator-beamformer produces 500 pulsar search beams with 118MHz of bandwidth and 16 pulsar timing search beams with up to 300MHz bandwidth each \cite{hampson_square_2022}.
The expected required latency for transient detection is 100$\mu$s \cite{aafreen_high-performance_2023} which is well above the minimum end-to-end response time of the Loihi 2 system \cite{davies_intel_nodate}.
In both cases however, implementing SNN logic in the existing FPGA resources is a sensible first step.
\subsection{Imaging}
Imaging is a data intensive HPC workload.
The main approach to turn visibilities into images, in essence, averages visibilities over time, and iteratively subtracts the brightest elements in the image, determined in Fourier space, until the known (or detected) bright foreground sources are removes and the system thermal limits are reached \cite{thompson_interferometry_2017, offringa_wsclean_2014}. 
This iteration of convolution and de-convolution is the main bottleneck, and while not ML based, large scale general purpose neuromorphic systems may one day provide a performant solution.
Incoming radio astronomy demands can consume vast computational resources, and therefore even a meager improvement in efficiency translates to massive operational benefit. 

Table \ref{tab:dep:imaging} contains estimates for the computational resources displaced and hypothetical, appropriately scaled neuromorphic resources that could replace them, if suitable techniques are researched. 
We single out the SpiNNaker 2 system as it contains both neuromorphic acceleration and more traditional ARM cores for general purpose compute, making such a system more applicable to general purpose computing tasks.

\begin{table}[!htbp]
\centering
\caption{Hypothetical neuromorphic deployments for radio astronomy imaging.}
\label{tab:dep:imaging}
\begin{tabularx}{\textwidth}{@{}cccccccc@{}}

\toprule

Instrument & Channels & Baselines & \begin{tabular}[c]{@{}c@{}}Proposed\\ Hardware\end{tabular} & Quantity & \begin{tabular}[c]{@{}c@{}}Current\\ Power (W)\end{tabular} & \begin{tabular}[c]{@{}c@{}}Proposed\\ Power (W)\end{tabular} & \multicolumn{1}{l}{\begin{tabular}[c]{@{}c@{}}Power\\ Proportion (\%)\end{tabular}} \\ \midrule

MWA        & 6400       & 8128        & \multirow{4}{*}{SpiNNaker 2}                                & 214      & 152000                                                     & 5136                                                         & 3.38                                                                                \\

ASKAP      & 16416      & 630          &                                                             & 17       & 152000                                                     & 408                                                          & 0.27                                                                                \\

LOFAR      & 124928     & 1326        &                                                             & 35       & 7280                                                       & 840                                                          & 11.54                                                                               \\

SKA-Low    & 65536      & 130816      &                                                             & 3443     & 597760                                                      & 82632                                                        & 13.82                                                                               \\ \bottomrule

\end{tabularx}
\begin{tablenotes}
    \item SKA-Mid and ngVLA have been omitted since no existing HPC resources for imaging exist. For SKA-Low, we have used the whole Setonix system at the Pawsey Supercomputing centre as reference. For the MWA and ASKAP we have used the GPU partition in isolation as reference.
\end{tablenotes}
\end{table}
We have calculated the number of boards required to process all channels and all baselines simultaneously, which is conservative, as some tolerance for buffering is typically permitted.
The most significant improvement comes for MWA and ASKAP whose minimally viable neuromorphic resources require 3.4\% and 0.3\% of the energy required by current resources.
LOFAR and SKA-Low both see an order of magnitude reduction in power consumption.

While speculative and preliminary, this modeling shows the potential for neuromorphic computing to provide significant practical benefits to radio astronomy.
\section{Discussion}\label{sec:discussion}
Radio astronomy could be a `killer application' for neuromorphic computing.
Under the lens of in-sensor computing, radio telescopes are among the largest and most data-intensive sensors constructed.
Radio observatories tie vast complex networks of dishes or antennas with serious computing resources.
Incoming telescopes, such as the SKA and ngVLA operate with sufficient scale to demand real-time processing of raw and correlated spectrographic data.
This demand, combined with a history of experimental computing methods, provides ample motivation to explore non-von Neumann and specifically neuromorphic, approaches to several data processing steps. 
We investigated several potential applications and localities for neuromorphic computing across modalities in existing and incoming radio telescopes, spanning flexible but near-term FPGA-based methods, medium-term commercial ASIC integration, and long-term large-scale neuromorphic systems.

Building on recent work into RFI detection with SNNs, we suggest that integrating real-time RFI detection into pre-correlated data is the most pragmatic and immediately viable approach.
Specifically, integrating SNN-based approaches into existing FPGA resources at the ASKAP, SKA-LOW and SKA-MID telescopes offers the most feasible near-term impact.
Under this approach, data-driven RFI detection methods could be integrated into several points in the signal processing chain without additional capital outlay.
Existing learning approaches have shown promise; however, biologically inspired de-noising, evolutionary algorithmic approaches and BPTT-trained anomaly detection approaches may yield more robust SNN-based RFI detection methods.
Moreover, if neuromorphic ASICs were to be included for RFI detection, such methods could use at least one but up to three orders of magnitude less energy than currently utilized resources.

Transient search is a long-suspected use case for neuromorphic computing and is likely the most scientifically impactful task to attempt beyond RFI detection.
While computational requirements, particularly with respect to latency, are more stringent, there is reason to believe that contemporary neuromorphic hardware could feasibly conduct end-to-end transient detection, though this remains for future work.

Imaging demands more general-purpose resources than the more SNN-suitable RFI detection and transient detection tasks.
However, under the assumption that neuromorphic hardware could perform imaging operations, we find reason to believe a similar 1-3 order of magnitude reduction in power consumption is possible. 
However, integrating neuromorphic computing at this stage in radio astronomy processing pipelines requires further research into general-purpose neuromorphic computing processors, alongside machine-learning and SNN-based techniques.

\begin{figure}
    \centering
    \includegraphics[width=\linewidth]{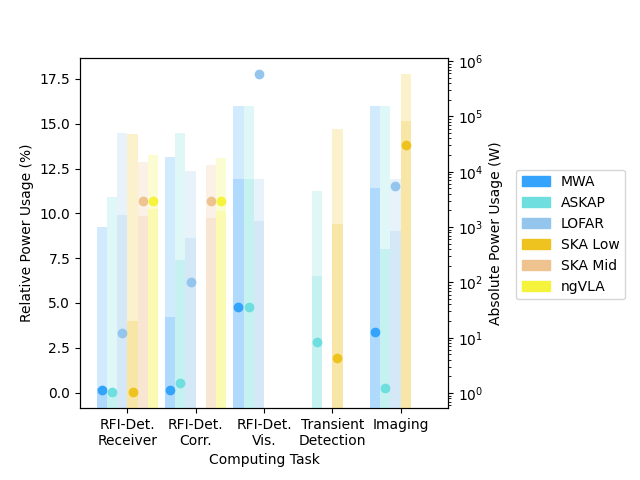}
    \caption[Predicted relative (circles, left axis) and absolute (bars, right axis, log-scale) power consumption for several radio astronomy processing tasks and several radio telescopes.]{Predicted relative (circles, left axis) and absolute (bars, right axis, log-scale) power consumption for several radio astronomy processing tasks and several radio telescopes. RFI detection at the receiving stage with purely neuromorphic computing tasks would consume up to three orders of magnitude less energy for MWA, ASKAP and SKA-Mid instruments. Correlator-scale RFI detection would provide almost the same relative improvement. Post-correlector RFI-detection provides around one order magnitude improvement, as would transient detection and imaging.}
    \label{fig:power_outlook}
\end{figure}

Figure \ref{fig:power_outlook} summarizes these potential gains, depicting the predicted relative and absolute power consumption for several instruments and several computing tasks, arranged in near to far-term feasibility.

While still maturing, the SNN and neuromorphic community is still burgeoning, and a continued focus in stable tooling and convincing benchmarks is necessary to drive support towards attempting large-scale data-processing challenges with neuromorphic computing.
Tooling that allows evaluation of SNN-based methods on FPGA resources that map cleanly into efficient ASIC implementations would be particularly useful in this case, where FPGA implementation is the most likely route to early adoption in the radio astronomy community.
Beyond any initial FPGA implementation, where the primary benefit is the enabling of data-driven processing tasks, the long-term inclusion of neuromorphic hardware would be transformative to telescope operations.
Moreover, demonstrating the ability for neuromorphic computing to handle large-scale spectrographic data processing, we hope to find other similar use-cases in related fields such as real-time medical imaging, RF signals intelligence, oceanography and seismic analysis, for example.

Finally, as telescopes have been traditionally constrained by their associated computing facilities, neuromorphic computing offers the even more distant, but exciting, possibility of supporting telescopes at least an order of magnitude larger than what is currently conceived.
The first steps towards neuromorphic astronomy have already been taken; the neuromorphs are on the march!
\section{Methods}\label{sec:methods}
\subsection{Processing Models}\label{subsec:processing_models}
We outline the processing models for each telescope under consideration, and our rationale for selecting various neuromorphic chipsets for various computing tasks.
For each, we provide an introduction to the telescope, its scientific goals, and then the processing chain from antenna to science archive.
Examples are provided in roughly construction chronological order and then by data intensity.
The signal chain, for brevity, has been simplified into a few key steps: Receiving, where raw voltage signals are digitized; Correlation, where signals from all antennas are combined; Beamforming, where signals from antenna are selectively combined to increase signal strength; Transient Detection, where rapidly changing phenomena are detected, often specifically for Pulsars, and Imaging, where correlated `visibility' data are averaged and iteratively refined into `data-cubes' for archival or downstream science. 
\subsubsection{MWA}
The Murchison Widefield Array (MWA) is a low-frequency radio interferometer telescope located at the Murchison Radio-astronomy Observatory in Western Australia, consisting of 256 phased array `tiles', each a 4x4 grid of dual-polarization dipoles, capturing signals across 70-300MHz \cite{wayth_phase_2018}.
The MWA was primarily built as a precursor to the Square Kilometre Array (SKA), the MWA aims to probe the Epoch of Reionization via 21 cm neutral hydrogen signals \cite{line_verifying_2024}, monitor solar and ionospheric activity \cite{tingay_murchison_2013}, detect radio transients \cite{hurley-walker_radio_2022}, and survey the extragalactic sky \cite{wayth_gleam_2015}. 

\begin{figure}
    \centering
    \includegraphics[width=\linewidth]{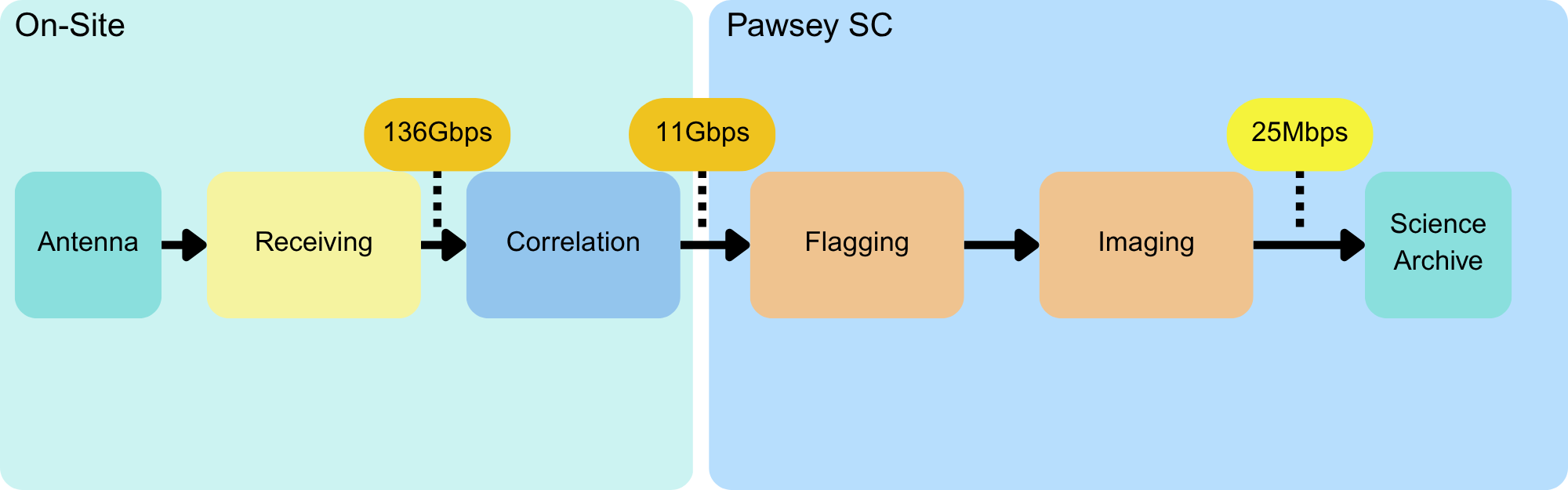}
    \caption{High-level data flow diagram for the MWA telescope. Cyan blocks represent data sources or sinks, yellow blocks represent FPGA processing, blue blocks represent GPU processing, and orange blocks represent CPU processing.}
    \label{fig:dataflow:MWA}
\end{figure}

Figure \ref{fig:dataflow:MWA} presents a high-level depiction of the MWA data-processing pipeline. 
Processing is split across three locations, on-site, where raw antenna signals are correlated into visibilities, transmitted $\approx$ 650km (403mi) via fibre optic cable to Curtin University in Perth, where they are buffered before post-correlation processing, imaging and pulsar searching happens, on the Setonix system at the Pawsey Supercomputing Centre.
The MWA digital receiver is primarily FPGA powered; 16 digital receivers process signals from 128 tiles (two observing modes use half of the 256 tiles each), and each digital receiver combines eight Xilinx Virtex-4 SX35 FPGAs, each consuming 5W of power, with a single Xilinx Virtex-5 SX50T consuming 3W of power \cite{prabu_digital-receiver_2015}.
Each Receiver receives two 8-bit wide streams at 655.36 mega samples per second \cite{prabu_digital-receiver_2015, girish_progression_2023}.
A total of 256 FPGA elements power the instrument with a combined power consumption of around 1024W.
The on-site correlator, MWAX, is GPU-powered \cite{morrison_mwax_2023} and allows for real-time operation, handling all active 128 station tiles.
The MWA correlator is powered by 24 GPU servers each armed with two AMD Epyc 7F72 CPUs and an Nvidia A40 GPU, each ingesting an average of 1.32GB/s \cite{morrison_mwax_2023}.
The Nvidia A40 consumes 300W \cite{noauthor_nvidia_nodate} and the AMD CPUs have a rated TDP of 240W \cite{advanced_micro_devices_amd_2020}, yielding a total power consumption of 18.72kW.
A key challenge with correlation is that, by definition, the signals from all antenna must be correlated with all other signals \cite{thompson_interferometry_2017}.
Each server ingests the data from a single coarse channel.
Each channel is comprised of 5-bit complex numbers segmented into packets representing 8 seconds of observation.
The input for the correlator engine ingests 6,400 signal channels and post-correlation data-rates comprise 6,400 fine-width channels (of 200MHz each), which can be averaged down from 6,400 to 1 channel depending on observation requirements.
Correlated signals are transmitted to Perth and handled by Setonix, comprised of 1592 AMD EPYC 7763 CPUs with a TDP of 280W and 304 AMD Instinct MI250X GPUs with a TDP of 500W each \cite{noauthor_setonix_nodate}.
This machine is shared with multiple science projects nationally; however, astronomy processing consumes a large proportion of available compute. 
The total visibility data rate is multiplied by the number of pairs of possible baselines, or pairs of antenna tiles, for the MWA.
This corresponds to 8,128 baselines requiring processing in parallel, but not in real-time.

Currently, post-correlation flagging and imaging-pre-processing are handled within the same Birli processing pipeline \cite{noauthor_mwatelescopebirli_2025} where flagging by AOFlagger \cite{offringa_aoflagger_2010} consumes $\approx$10\% processing time for an observation.
Transient detection is primarily GPU-based, while imaging steps use a mix of CPU and GPU resources.

In considering RFI detection with neuromorphic resources for the MWA, we envisage embedding 16 Xylo 2-like processors (each handling 8 channels of input) per receiver element, requiring 2048 chips with a combined energy usage of 1.23W, or 0.1\% of the existing resources.
RFI detection at the correlator needs to handle all 6,400 input channels.
Given existing techniques for RFI detection with SNNs use \textless5000 neurons, a single Intel Loihi 2 Oheo Gulch chip matched for each existing GPU resource should suffice, consuming 24W total or 0.1\% of existing resources.
For post-correlator RFI-detection, we look to replace the GPU partition of the Setonix supercomputer which runs many MWA imaging processes with SpiNNaker2 boards.
Even this overly conservative estimate would require 7.3kW or 5\% of the existing resources required to run the existing machine.
The MWA is a good first example to present, combining some of the major challenges in signal processing and transport close to the antennas, but with a (slightly) more forgiving processing pipeline downstream, compared to some of the more data-intensive instruments we introduce later.
\FloatBarrier
\subsubsection{ASKAP}
The Australian Square Kilometre ARray Pathfinder (ASKAP) is a dish-based radio interferometer also located at the CSIRO Murchison Radio-astronomy Observatory in Western Australia.
ASKAP comprises of 36 12-meter parabolic antennas equipped with phased-array feeds (PAFs) each consisting of 188 receiving elements each.
The instrument operations in the 700-1800MHz range with construction commencing in 2009, completed in 2012 with pilot surveys and several upgrades until full operations commenced in 2022 \cite{hotan_australian_2021}.
ASKAP is a technology demonstrator to the Square Kilometre Array (SKA), and as such focuses on rapid wide-field surveys \cite{duchesne_rapid_2025}, extragalactic HI and continuum observations \cite{pingel_gaskap-hi_2022}, and characterizing radio transients and pulsars \cite{hobbs_pilot_2016}. 

\begin{figure}
    \centering
    \includegraphics[width=\linewidth]{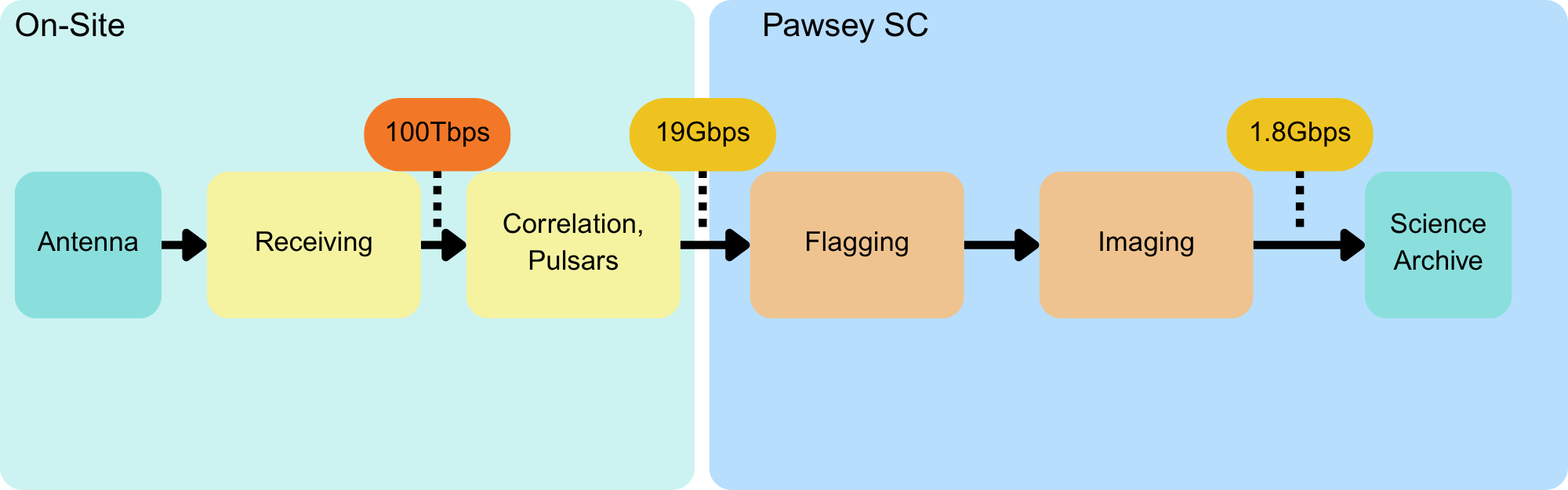}
    \caption{High-level data flow diagram for the ASKAP telescope. Diagram conventions from Figure \ref{fig:dataflow:MWA} apply.}
    \label{fig:dataflow:ASKAP}
\end{figure}

Figure \ref{fig:dataflow:ASKAP} presents a graphical depiction of the data-processing workflow for ASKAP.
FPGAs handle receiver processing.
On each antenna, 12 `Dragonfly' modules handle immediate analog to digital conversion with 4 Xilinx Kintex 7 325T FPGAs each with 12-bit resolution \cite{brown_design_2014}.

This hardware handles sampling clocks of 1280 and 1536MHz producing either 640 or 768 1MHz (width) signal channels represented by 16-bit complex value streams.
In total, each antenna produces 192 optical signals, comprised of 188 optical signals, 2 calibration signals and 2 reserved for future use such as RFI-mitigation \cite{hotan_australian_2021}.
This is where we would include neuromorphic resources for RFI detection, requiring 48 Xylo-2 chips per antenna, or 1728 in total, for 1W of power.
In lieu of measured power consumption, we estimate the power use of these FPGA at 2W each, which across all 36 antennas corresponds to 1728 devices and 3.64kW of power total.
The output of the digital receiver for each antenna is correlated and beamformed by seven `Redback' modules, each comprised of 6 Xilinx Kintex 7 XC7K480T FPGAs, each receiving 48MHz of bandwidth from each antenna from all 192 digital ports \cite{hotan_australian_2021}.
Each FPGA always produces 423 frequency channels sampled at 1.185 MHz.
Each `Redback' module has a maximum power consumption of 373W and nominal consumption of 200W \cite{hampson_askap_2014}.
There are 252 `Redback' modules at ASKAP (seven for each of the 36 dishes), bringing a total of 1,512 FPGAs.
To approximate power usage, we split 200W across six FPGAs for 33W each and total power consumption is estimated at 49.90kW.
In considering sufficient neuromorphic resources at the correlator stage, we need to handle the 423 frequency channels input to each `Redback' module.
Given the existing scale of SNN-based RFI detection methods, again, a Loihi 2 Oheo Gulch chip for each of the 252 `Redback' modules yielding a power-consumption of 252W or 0.5\% of the existing power budget.

The correlator outputs two signal streams, long-term accumulated visibilities which are downlinked to the Pawsey Supercomputing Centre for imaging and beamformed signals for transient (pulsar) detection by the Commensal Realtime ASKAP Fast Transient COherent (CRACO) system \cite{wang_craft_2025}.
`CRACO' is comprised of 20 Xilinx Alveo U280 high-performance FPGAs where RFI detection, and transient searching is run.
`CRACO' is aiming to generate $256 \times 256$ images with a goal of 1.7ms time resolution, yielding 23 Terapixels per second of data \cite{wang_craft_2025}.
We estimate power consumption for each chip at 225W \cite{advanced_micro_devices_alveo_2023} bringing system power consumption estimates to 4.5kW.
For neuromorphic processing at the transient detection stage, FPGA processing is so integral, it is unlikely that integrating a different class of processor is feasible without replacing the entire transient search process.

Downstream visibility processing, beginning with RFI flagging and subsequently imaging, containing up to 16,416 frequency channels and 630 baselines \cite{hotan_australian_2021} is also handled at the Pawsey center with a bespoke processing suite, ASKAPsoft \cite{guzman_askap_2019}, and is shared with MWA and other science projects.
Processing is handled in pseudo-real time, handling 2.4GBps of incoming data, processing each observation after completion \cite{guzman_are_nodate}.
An 8-hour observation will produce $\approx$70TB of data \cite{whiting_high-performance_2018} before archiving the resulting data products to long-term storage.
For neuromorphic resources, the MWA and ASKAP share the same super-computing facilities, and therefore the same post-correlator estimates apply.

\FloatBarrier
\subsubsection{LOFAR}
The Low-Frequency Array (LOFAR) is a large-scale radio interferometer telescope headquartered in the Netherlands with further stations across Europe and consists of around 100,000 dipole antennas grouped into 52 stations \cite{van_haarlem_lofar_2013}.
The instrument is divided into 24 core stations in the Netherlands, 14 remote stations distributed over the Netherlands, and 14 international stations distributed across Europe.
LOFAR was conceived as a precursor instrument to the full SKA while making significant contributions in its own right.
Construction began in 2006 with full operations commencing in 2012.
LOFAR 2.0 upgrades started in 2021 and are ongoing through 2025.
LOFAR is designed primarily to explore the early Universe, conduct large-scale surveys and detect transient events.

\begin{figure}
    \centering
    \includegraphics[width=\linewidth]{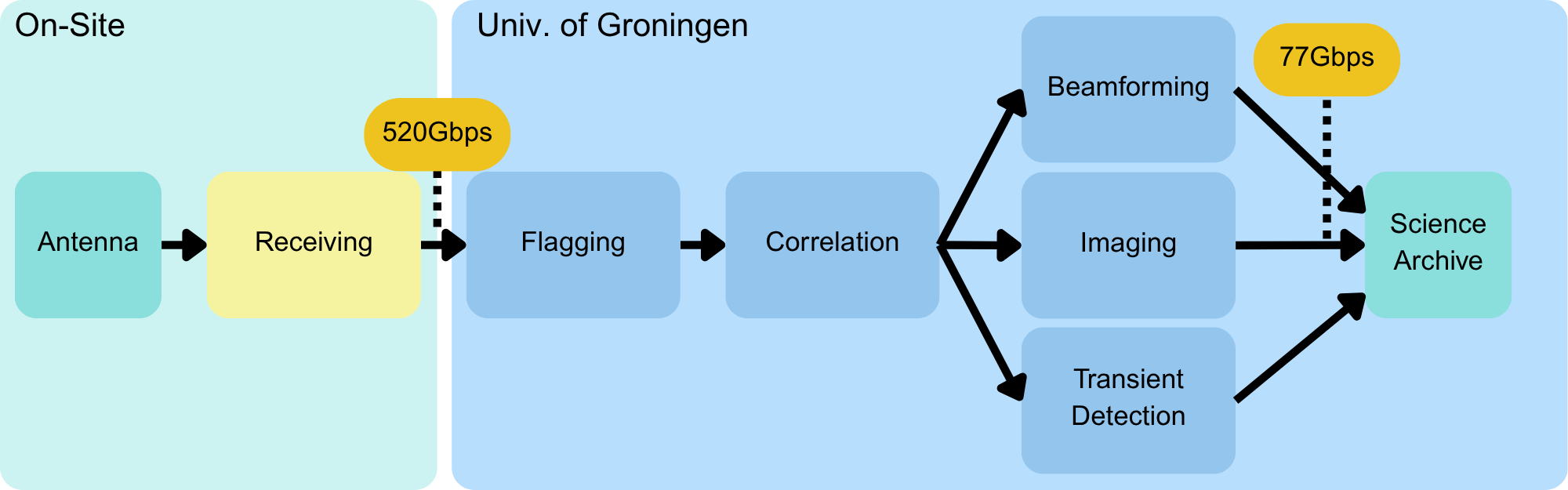}
    \caption{High-level data flow diagram for the LOFAR telescope. Diagram conventions from Figure \ref{fig:dataflow:MWA} apply.}
    \label{fig:dataflow:LOFAR}
\end{figure}

Figure \ref{fig:dataflow:LOFAR} depicts the data processing workflow for the LOFAR telescope.
Data processing involves digitizing signals at each station before transmission to a central correlator in Groningen for real-time correlation before further imaging, or beam-forming for pulsar and transient searching.
Petabyte-scale archives are managed in Amsterdam and distributed globally using the LOFAR Long Term Archive.

The computing for each station is provided by 8 Uniboard\textsuperscript{2} devices \cite{schoonderbeek_uniboard2_2019, juerges_lofar20_2022} which contain 4 Intel Arria A10GX115 FPGAs yielding 1664 devices which we estimate use $\approx$30W, resulting in 49.9kW power consumption across the whole instrument.
The per-station processing handles beamforming and correlation for each station, before transmitting the 3-9Gbps data stream \cite{juerges_lofar20_2022} via 10Gbps Ethernet to the central processing system in Groningen.
Therefore, as the per-station data rate is more intense than previously discussed instruments, we suggest a Loihi 2 Oheo Gulch chip per station Uniboard would suffice.
1664 chips in total would be required at 1664W power consumption, or 3\% of the existing power requirement.
 
In Groningen, the GPU-powered COBOLT correlator \cite{broekema_cobalt_2018}, which has been recently upgraded \cite{juerges_lofar20_2022} handles final correlation and downstream processing, including post-correlation flagging and imaging, which are performed by a bespoke cluster.
COBOLT consists of 13 computing nodes containing two Nvidia Tesla V100 GPUs with a TDP of 250W \cite{nvidia_corporation_nvidia_2020} each and two Intel Xeon Gold SP6140 CPUs with a TDP of 140W \cite{intel_intel_2017}.
The total power consumption for this cluster stands around 10kW.
There is particular interest in applying SNN-based RFI detection to LOFAR, as existing work has already attempted RFI detection on LOFAR datasets \cite{pritchard_spiking_2024}.
However, the telescope just recently received a GPU-centric upgrade and therefore is considered a long-term prospect. Nonetheless, at the correlator, a higher bandwidth, more flexible platform would suffice, and we therefore consider a SpiNNaker2 board per existing GPU, of which there are 26, requiring 624W in total power, or 6\% of the existing power usage.
The science data processor is a separate cluster consisting of 50 CPU nodes equipped with Intel Xeon E5-2680v3 with 120W TDP \cite{intel_intel_2014} and four GPU nodes equipped with an Intel Xeon E5-2630v3 (85W TDP \cite{intel_intel_2014_2}) and Nvidia K20X (235W TDP \cite{nvidia_corporation_nvidia_2013}) with a combined TDP of 7.3kW.
This machine needs to handle up to 124,928 frequency channels and 1,326 baselines \cite{van_haarlem_lofar_2013}.
RFI flagging and imaging are handled by custom software packages such as AOFlagger \cite{offringa_aoflagger_2010} and WSClean \cite{offringa_wsclean_2014}, where a net data output of 77Gbps is archived for long-term storage and downstream science use.
At the post-correlator stage, a more general-purpose neuromorphic machine matches the batch-processed style of computing at the imaging stage.
There are 2048 input channels which is well within the capability of several neuromorphic chipsets, but we consider 54 SpiNNaker2 boards with a combined power consumption of 1.3kW or 18\% of the original power budget.

LOFAR represents a close analogue to the SKA-Low instrument, albeit with an order of magnitude lower data-rate to deal with.
There is interest in integrating neuromorphic computing methods into the instrument's processing pipeline \cite{bos_analogue_2025}, although this work is still preliminary. 
\FloatBarrier
\subsubsection{SKA-LOW}
The Square Kilometer Array (SKA) is a global mega-science project to build the world’s largest and most sensitive radio telescope, split into SKA-Low in Australia and SKA-Mid in South Africa, designed to explore the Universe’s early history, galaxy evolution, cosmic magnetism, and fundamental physics, with construction underway and full operations targeted for the early 2030s \cite{schilizzi_evolution_2024}.
The SKA is the result of decades of planning, design and construction realized as a collaboration between 35 member countries, and is already a storied project \cite{schilizzi_evolution_2024}

The low-frequency instrument is also situated at the CSIRO Murchison Radio-astronomy Observatory in Western Australia.
When fully constructed, SKA-LOW will comprise 131,072 dual-polarization dipole antennas clustered into 512 stations with 256 antennas each, and operations across the 50-350MHz range \cite{braun_advancing_2015}.
Construction officially commenced in 2021 with first light in 2025 and completion expected around 2030 \cite{schilizzi_evolution_2024}.
SKA-LOW is focused on the Universe's first billion years along side transient events.
SKA-Low is designed to offer 25\% better resolution, eight times the sensitivity and 135 times faster survey speeds than predecessor instruments like LOFAR \cite{braun_advancing_2015}.

\begin{figure}
    \centering
    \includegraphics[width=\linewidth]{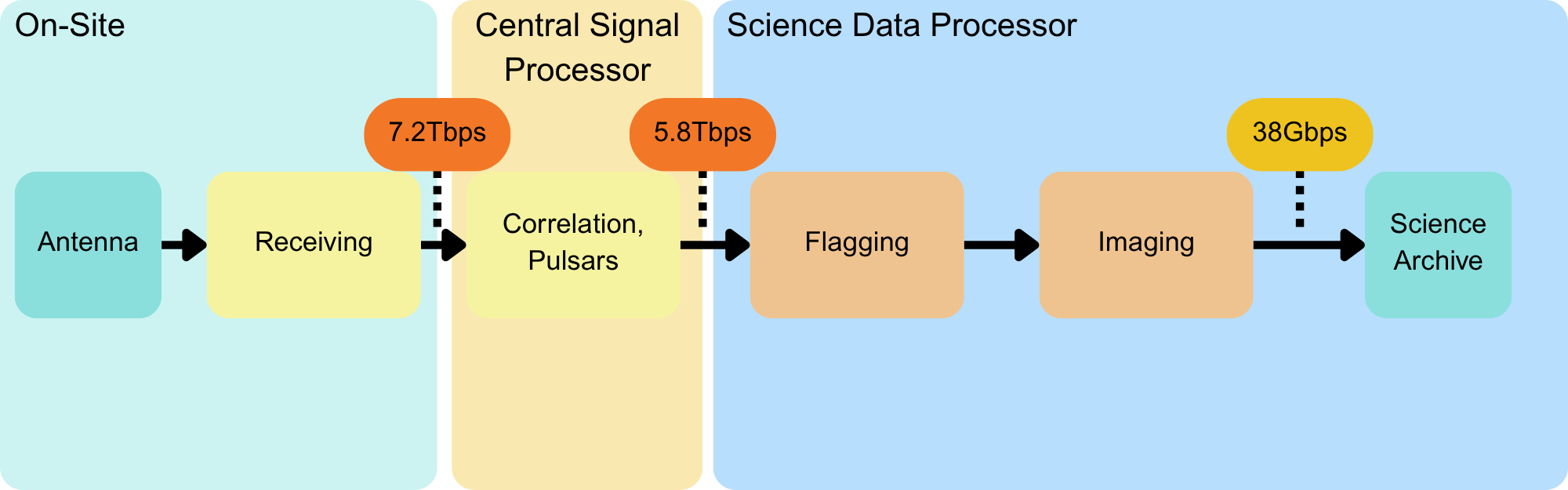}
    \caption{High-level data flow diagram for the SKA-Low telescope. Diagram conventions from Figure \ref{fig:dataflow:MWA} apply.}
    \label{fig:dataflow:SKA-LOW}
\end{figure}

Data processing for SKA-Low correspondingly represents an order of magnitude increase in volume.
Figure \ref{fig:dataflow:SKA-LOW} depicts the data flow for the SKA-Low instrument. 
Raw signals are digitized at the stations and collectively transmitted at 7.2 Terabits per second (Tb/s) to an on-site Central Signal Processor (CSP) for cleaning and averaging.
Each station is handled by eight `tile-processor modules' TPMs \cite{chaudhari_station_2024}, each consisting of 2 Xilinx XCKU040 FPGAs \cite{benthem_aperture_2021}, which we estimate consume $\approx$6W each, requiring 49kW total for all 512 stations.
Each station produces 512 beams, which are correlated downstream.
Deploying neuromorphic chipsets alongside the existing FPGA resources per-station is intuitive, producing 512 channels each. We recommend 64 SynSense Xylo 2 chips per station, bringing a combined total of 32,768 chips with a combined power usage of 19.7W or 0.1\% of the total power consumption.
The correlator for SKA-Low also handles pulsar searching and is comprised of 400 high-bandwidth-memory equipped Xilinx Alveo U55C FPGAs, and is termed `atomic COTS' \cite{hampson_square_2022}.
This cluster consumes 6Tbps data but produces 9Tbps of output and consumes 60kW in total.
This machine also handles RFI flagging implemented in FPGA logic, utilizes \textless50\% of the available FPGA logic; with the FPGAs chosen to have headroom for further developments \cite{hampson_square_2022}. 
Atomic COTS produces 16 Pulsar timing beams, 500 Pulsar search beams and 2.7Tbps of visibilities which are transmitted to Perth to the Science Data Processor (SDP) comprised of typically 55,296 channels of 5.4 kHz width with an integration time of 0.85s.
Considering correlation is handled by significant FPGA resources, including neuromorphic ASICs without replacing the entire process is infeasible and therefore omitted.

Then, the data is transmitted at a rate of up to 5.76 Tb/s to Perth where a dedicated HPC system handles workflows to generate science-ready data products.
SKA-Low can produce up to 65,536 frequency channels and visibilities for 130,816 baselines \cite{hampson_square_2022}.
The SDP does not yet exist, so for the purpose of pre-construction commissioning, various HPC resources, including the Pawsey Supercomputer center have been utilized.
The requirements for the SDP are intense \cite{wang_processing_2020} and handling I/O for imaging has long been known as the largest data-processing challenge of all, although innovating techniques may close the gap \cite{williamson_optimising_2024}.
Final products are exported to an international network of SKA Regional Centers for remote astronomer access.
\FloatBarrier
\subsubsection{SKA-MID}
The SKA Mid-frequency (SKA-Mid) telescope is comprised of 197 dish antennas, located in the Karoo desert of South Africa's Northern Cape \cite{swart_highlights_2022}.
The dishes operate across 350MHz to 15.4GHz.
Construction began in 2021 with first light in 2025 with an expected completion time by 2030.
SKA-Mid is designed to study galaxy evolution, neutral hydrogen, and transient phenomena, which have been prototyped by the MeerKAT instrument \cite{jonas_meerkat_2018}, which will be absorbed into the SKA-Mid instrument \cite{braun_advancing_2015}.
Together the two SKA instruments will perform observations in collaboration, but also separately, being jointly operated from a headquarters in Manchester, UK.

\begin{figure}
    \centering
    \includegraphics[width=\linewidth]{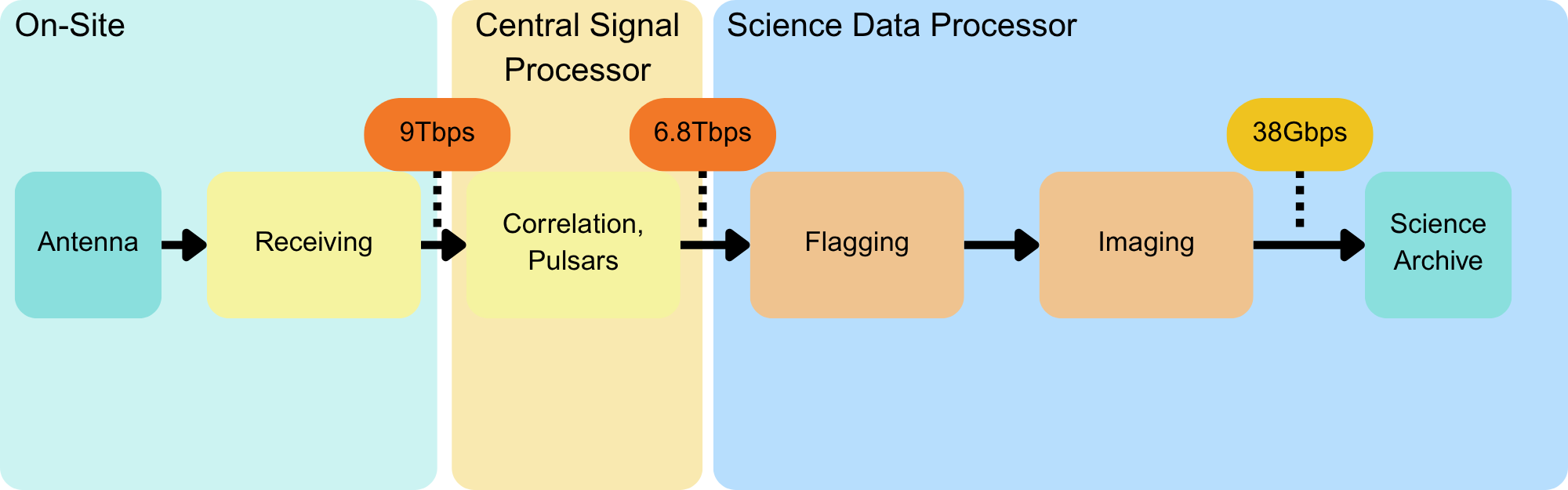}
    \caption{High-level data flow diagram for the SKA-Mid telescope. Diagram conventions from Figure \ref{fig:dataflow:MWA} apply.}
    \label{fig:dataflow:SKA-MID}
\end{figure}

Data processing at a high level is similar to the SKA-Low, albeit with different data shaping.
Figure \ref{fig:dataflow:SKA-MID} depicts the data flow for the SKA-Mid instrument. 
FPGA resources digitize signals at each dish before transmission to an on-site Central Signal Processor (CSP) for correlation, beamforming and pulsar searching, then processing at the SKA regional center in Cape Town on HPC resources before exporting science-ready data products to the wider SRC-net.
Digitization is handled on the dish pedestal \cite{roy_overview_2022, hall_innovative_2016}, with initial processing handled by an Altera Stratix 10 DX FPGA, which we estimate uses 75W of power, followed by correlation handled by a centralized TALON DX correlator \cite{pleasance_talon_2021} powered, again, by an Altera Stratix 10 DX FPGA, which we estimate uses 75W of power, and 180 such boards will be required for the full SKA-Mid deployment.
For suggesting neuromorphic hardware at the receiver stage for RFI-detection, we consider simultaneous deployment of an Intel Loihi 2 Kapoho point per dish, consuming 8W each for a combined total of 1576W over the 197 dishes or 11\% of the existing budget.
At the correlator stage, we again consider an Intel Loihi 2 Kapoho point, consuming 1440W and 11\% over the 180 units required.
Like SKA-Low, SKA-Mid will produce up to 65,536 frequency channels and 19,306 baselines for imaging purposes \cite{swart_highlights_2022}.
Total power consumption for both machines is around 28.2kW.
Downstream image processing resources, like with SKA-Low, have not yet been provisioned, and hence, we do not provide reference hardware nor neuromorphic suggestions.

While both SKA-Low and SKA-Mid share many design motifs, the differing antenna modalities and structure of data processing place different strains on the computing resources available. 
\FloatBarrier
\subsubsection{ngVLA}
The Next Generation Very Large Array (ngVLA) is a planned radio interferometer telescope primarily based in New Mexico, USA, with other stations across North America, comprising of 263 fixed 18-meter dishes and 19 movable 6-meter dishes operating across 1.2-116GHz \cite{selina_next-generation_2018}.
Initiated by the National Radio Astronomy Observatory (NRAO), design and prototyping began in 2015, with construction approved in 2024 and major site work starting in 2025; first science operations are targeted for 2031, with full completion by 2037.
The ngVLA aims to study planet and star formation, astro-chemistry, galaxy evolution, and relativistic phenomena like black holes with 10 times the sensitivity and resolution of the current VLA and ALMA \cite{selina_next-generation_2018}.

\begin{figure}
    \centering
    \includegraphics[width=\linewidth]{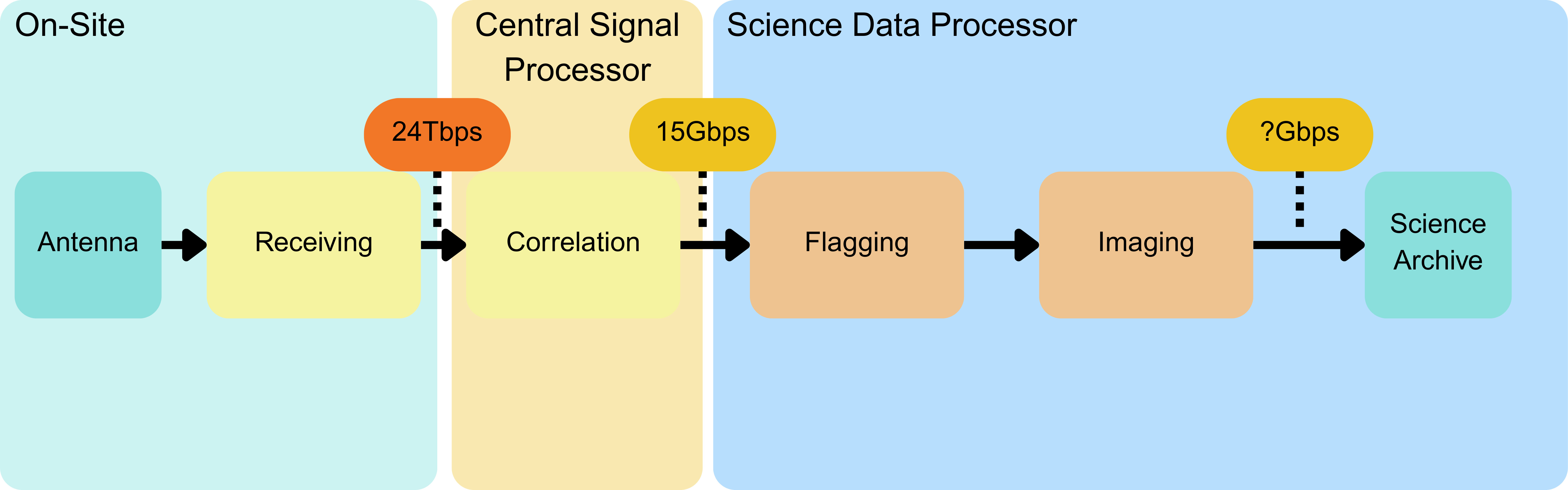}
    \caption{High-level data flow diagram for the ngVLA telescope. Diagram conventions from Figure \ref{fig:dataflow:MWA} apply.}
    \label{fig:dataflow:ngVLA}
\end{figure}

Figure \ref{fig:dataflow:ngVLA} shows the proposed data flow for the ngVLA instrument.
Data processing involves digitizing signals at each antenna, transmitting typically at 720Gbps to a central correlator in Socorro, New Mexico, for real-time calibration and imaging \cite{selina_ngvla_2024}, with pipelines producing science-ready datasets of several petabytes annually, archived and distributed through NRAO’s data centers for global access.
While construction has yet to commence, we can still reason about some preliminary computing requirements.
Pedestal processing looks to follow a similar design to the TALON correlator developed for SKA-Mid \cite{pleasance_talon_2021} and therefore we assume the use of Altera Stratix 10 DX FPGAs for receiving, correlation, and pulsar searching, with a similar proportion to the SKA-MID.
Therefore, 263 FPGAs for the dishes and a further 240 for the correlator, consuming 19.7kW and 18kW of energy respectively.
For roughly equivalent neuromorphic resources, for each dish, an Intel Loihi 2 Kapoho point chip at the receiver and correlator stage, which for 263 dishes and 240 correlator boards will require 2.1kW and 1.9kW, respectively, or 11\% of the original power consumption in both cases.
ngVLA will produce visibilities with up to 300,000 frequency channels, the most of any planned or existing instrument, with 34,453 baselines \cite{selina_next-generation_2018}.
Again, imaging resources have not yet been considered, and we therefore refrain from commenting further. 
Design of the ngVLA is still ongoing, and therefore represents the most contemporary thinking about dish-based radio telescope design, hallmarked by near ubiquitous use of FPGA resources, owing to the flexibility and latency requirements of radio astronomy.

\subsubsection{Existing Computer Hardware}\label{subsubsec:existing_hardware}

We first discuss the existing resources at various telescope facilities, summarized in Table \ref{tab:impact:meth:traditional}.
Care has been taken, particularly with respect to FPGA resources, to provide an indicative power consumption estimate and overall resource characterization where possible.

\begin{table}[!htbp]
\centering 
\caption{Traditional hardware parameters.}
\label{tab:impact:meth:traditional}
\begin{tabularx}{\textwidth}{@{}ccccccc@{}}
\toprule
Device                       & Type & \makecell{Power\\(W)} & Instrument & \makecell{Processing\\Stage}        & Number & \makecell{Total Power\\(W)} \\ \midrule
\makecell{Xilinx Virtex 4 SX35}         & FPGA & 5         & MWA        & Receiving               & 128    & 640             \\
\makecell{Xilinx Virtex 5 SX50T}        & FPGA & 3         & MWA        & Receiving               & 128    & 384             \\
\makecell{Nvidia A40}                   & GPU  & 300       & MWA        & Correlation             & 24     & 7200           \\
\makecell{AMD Epyc 7F72}                & CPU  & 240       & MWA        & Correlation             & 48     & 11520          \\
\makecell{AMD EPYC 7763}                & CPU  & 280       & \makecell{MWA,\\ASKAP} & Imaging                 & 1592   & 445760         \\
\makecell{AMD Instinct MI250X}          & GPU  & 500       & \makecell{MWA,\\ASKAP} & \makecell{Imaging,\\Transients}     & 304    & 152000         \\
\makecell{Xilinx Kintex 7 325T}         & FPGA & 2*        & ASKAP      & Receiving               & 1728   & 3456           \\
\makecell{Xilinx Kintex 7\\XC7K480T}     & FPGA & 33        & ASKAP      & Correlation             & 1512   & 49896          \\
Xilinx Alveo U280            & FPGA & 225       & ASKAP      & Transients              & 20     & 4500           \\
Intel Arria A10GX115         & FPGA & 30        & LOFAR      & Receiving               & 1664   & 49920          \\
Intel Xeon Gold SP6140       & CPU  & 140       & LOFAR      & Correlation             & 26     & 3640           \\
Nvidia Tesla V100            & GPU  & 250       & LOFAR      & Correlation             & 26     & 6500           \\
Intel Xeon E5-2680v3 & CPU  & 120       & LOFAR      & Imaging                 & 50     & 6000           \\
Nvidia K20X                  & GPU  & 235       & LOFAR      & Imaging                 & 4      & 940             \\
Intel Xeon E5-2630v3 & CPU  & 85        & LOFAR      & Imaging                 & 4      & 340             \\
Xilinx XCKU040               & FPGA & 6*        & SKA-Low    & Receiving               & 8192   & 49152          \\
Xilinx Alveo U55C            & FPGA & 150       & SKA-Low    & \makecell{Correlation,\\Transients} & 400    & 60000          \\
Altera Stratix 10 DX         & FPGA & 75*       & SKA-Mid    & Receiving               & 197    & 14775          \\
Altera Stratix 10 DX         & FPGA & 75*       & SKA-Mid    & \makecell{Correlation,\\Transients} & 180    & 13500          \\
Altera Stratix 10 DX         & FPGA & 75*       & ngVLA      & Receiving               & 263    & 19725          \\
Altera Stratix 10 DX         & FPGA & 75*       & ngVLA      & \makecell{Correlation,\\Transients} & 240    & 18000          \\ \bottomrule
\end{tabularx}
\begin{tablenotes}
    \item Total compute resources are summarized from several sources and for several telescopes. IO Bandwidth and power provided where possible for FPGA resources. Total resources for the Setonix supercomputer at Pawsey Supercomputing Centre, which are shared across many science projects. Imaging resources for SKA-Low/Mid are not provided as they have not been decided yet.
    \item \textsuperscript{*}Power consumption estimated.
\end{tablenotes}
\end{table}
\subsection{Neuromorphic Hardware Specifications}\label{subsec:neuro_hw}
We now discuss the neuromorphic chipsets under consideration.
We choose to focus on a few select options, SynSense Xylo 2, Intel Loihi 2 and SpiNNaker 2, based on the ability to find concrete specifications and feasible hardware delivery.
Our pragmatic selection focuses on chips with relatively mature development ecosystems.
Table \ref{tab:meth:neuromorph} contains relevant specifications used to model hypothetical deployment in radio astronomy circumstances, along-side crude power consumption measurements.
Some values are estimates, in lieu of concrete, verified metrics, but have been calculated to provide an indicative performance or scale estimate, in particular with respect to I/O and power consumption specifications.

\subsubsection{SynSense Xylo 2}
The SynSense Xylo 2 platform is designed primarily for edge-sensing problems, characterized by extremely low power consumption, but a limited network size.
Each chip can support 16 inputs and 8 outputs and up to 1000 internal neurons \cite{noauthor_xylo-audio_2022} while drawing power consumption around 600mW for the task of RFI detection in particular for 512 frequency channels \cite{pritchard_neuromorphic_2025}.

\subsubsection{SpiNNaker2}
SpiNNaker 2 is the second-generation SpiNNaker chip \cite{gonzalez_spinnaker2_2024} coupling general purpose ARM cores with numerical accelerators, and contains 152 cores per chip, split between 38 quad-core processing elements (PE), each PE is capable of simulating between 250 to 1000 neurons depending on connection density, and a board consists of 48 chips \cite{spinncloud_spinnaker2_2025} connected in a two-dimensional toroidal mesh \cite{young_review_2019}.
The power consumption of the SpiNNaker2 system is a conservative estimate based on the language modeling system benchmark of 390mW \cite{nazeer_language_2024}.
The I/O constraints are derived from a per-processing-element limit of 500 incoming spikes per time step and a 10Gbps limit per board.

\subsubsection{Intel Loihi 2}
Intel's second-generation neuromorphic system, Loihi 2 implements SNNs with programmable dynamics \cite{orchard_efficient_2021}.
Each chip integrates 128 neuromorphic cores to support 1 million neurons in an asynchronous design.
Loihi 2's modular connectivity allowing planar connections between individual chipsets \cite{young_review_2019} permits several form factors combining multiple chips ranging from a single chip board (Oheo Gulch) to a HPC-scale Hala-point system \cite{abreu_neuromorphic_2025}.
Each chip comes with a 10Gbps Ethernet link, which provides an upper limit on the incoming information.
Power consumption is based on worst-case metrics scaled according to the number of chips available in each system \cite{shrestha_efficient_2024, brehove_sigma-delta_2025}.

\begin{table}[!htbp]
\caption{Neuromorphic hardware parameters}
\label{tab:meth:neuromorph}
\begin{tabular}{@{}llll@{}}
\toprule
Device                     & Neuron Count          & Clock-speed (MHz) & Power Consumption (mW) \\ \midrule
SynSense Xylo 2            & 1000                  & 6.25 - 100        & 0.6                    \\
SpiNNaker 2 Chip           & 32,250 - 152,000      & 100 - 400         & 500                    \\
SpiNNaker 2 Board          & 1,500,000 - 7,296,000 & 100 - 400         & 24000                  \\
Intel Loihi 2 Oheo Gulch   & 1,000,000             & \textgreater{}5*  & 1000                   \\
Intel Loihi 2 Kapoho point & 8,000,000             & \textgreater{}5*  & 8000                   \\
Intel Loihi 2 VPX          & 16,000,000            & \textgreater{}5*  & 16000                  \\
Intel Loihi 2 Alia Point   & 128,000,000           & \textgreater{}5*  & 128000                 \\
Intel Loihi 2 Hala Point   & 1,152,000,000         & \textgreater{}5*  & 1152000                \\ \bottomrule
\end{tabular}
\begin{tablenotes}
    \item Neuromorphic hardware parameters. The neuron count of a neuromorphic system is of clear relevance to SNN-based computing tasks. Clock-speed is relevant to latency-sensitive tasks such as transient detection and power consumption is used to generate instrument-scaled estimates of power usage.
    \item SynSense Xylo values have come from the official spec-sheet \cite{noauthor_xylo-audio_2022} Intel Loihi 2 power consumption values have been estimated from scaling up the power draw of a single Oheo Gulch chip in the worst case. SpiNNaker 2 I/O values have similarly been estimated from a rounded and scaled estimate from a taxing language modeling example \cite{nazeer_language_2024}.
    \item \textsuperscript{*}an asynchronous design makes clock-speed difficult to determine, but a minimum chip-wide time step under 200ns gives a lower-limit on effective clock-speed \cite{davies_intel_nodate}.
\end{tablenotes}
\end{table}
\FloatBarrier
\backmatter
\bmhead{Funding Declaration}
This work was supported by a Westpac Future Leaders Scholarship, an Australian Government Research Training Program Fees Offset and an Australian Government Research Training Program Stipend.

\bmhead{Competing Interests}
The authors have no competing interests to declare.

\bmhead{Author Contributions}
N.J.P conceived of the paper, prepared the figures and wrote the main manuscript text. R.D and A.W assisted in preparing the analysis. All authors reviewed the manuscript.

\newpage
\renewcommand{\tablename}{Supplementary Table}
\renewcommand{\figurename}{Supplementary Figure}
\setcounter{table}{0}
\section*{Supplementary Information}

\maketitle
\subsection*{Supplementary Note 1: SNN-based RFI detection results on the synthetic HERA dataset}

\begin{table}[!htbp]
\centering
\caption{Detection results for various algorithmic, ANN and SNN-based approaches to RFI detection on a synthetic HERA dataset.}
\label{tab:res:hera}
\begin{tabularx}{\linewidth}{@{}cccccc@{}}
\toprule
Work                & Model                & AUROC & AUPRC & F1    & Num-Parameters \\ \midrule
\multicolumn{6}{l}{\textit{Algorithmic Baselines}}                                           \\
\citet{mesarcik_learning_2022}            & AOFlagger \cite{offringa_aoflagger_2010}            & 0.974 & 0.88  & 0.873 & N/A            \\ \midrule
\multicolumn{6}{l}{\textit{ANN Baselines}}                                                   \\
\citet{mesarcik_learning_2022}                & R-Net \cite{sadr_deep_2020}                & 0.975 & 0.846 & 0.846 & 32.9k          \\
\citet{mesarcik_learning_2022}              & U-Net \cite{akeret_radio_2017}                & 0.975 & 0.896 & 0.902 & 292k           \\
\citet{mesarcik_learning_2022}            & AutoEncoder          & 0.981 & 0.927 & 0.91  & 365k               \\
\citet{vanzyl_remove_2024} & RFDL                 & 0.994 & 0.965 & 0.944 & 234k               \\
\citet{dutoit_comparison_2024}              & ASPP                 & -     & -     & 0.985 & 292k           \\
\citet{dutoit_comparison_2024}              & RNet-7 \cite{sadr_deep_2020}               & -     & -     & 0.989 & 32.9k          \\
\citet{dutoit_comparison_2024}              & RFI-Net \cite{yang_deep_2020}              & -     & -     & 0.993 & 809k           \\ \midrule
\multicolumn{6}{l}{\textit{SNN Baselines}}                                                   \\
\citet{pritchard_rfi_2024}           & ANN2SNN              & 0.944 & 0.92  & 0.953 & 339k               \\
\citet{pritchard_supervised_2024}           & BPTT                 & 0.929 & 0.785 & 0.761 & 73.7               \\
\citet{pritchard_spiking_2024}           & BPTT + DN            & 0.996 & 0.914 & 0.907 &  213k              \\
\citet{pritchard_advancing_2025}        & Liquid State Machine & 0.842 & 0.781 & 0.743 &  8.2k              \\
\citet{pritchard_polarisation-inclusive_2025}           & Full Polar           & 0.997 & 0.96  & 0.955 & 51.7k                \\
\citet{pritchard_neuromorphic_2025}           & BPTT-64              & 0.988 & 0.983 & 0.983 & 368k               \\ \bottomrule
\end{tabularx}
\begin{tablenotes}
    \item SNN-based detection methods are among the smallest models while performing very close to the state of the art ANN-based models.
    \item The HERA dataset is available freely online \cite{mesarcik_learning_2022}.
\end{tablenotes}
\end{table}
\newpage

\subsection*{Supplementary Note 2: SNN-based RFI detection results on the real LOFAR dataset}
\begin{table}[!htbp]
\caption{Detection results for various algorithmic, ANN and SNN-based approaches to RFI detection on a real LOFAR dataset.}
\label{tab:res:lofar}
\begin{tabularx}{\linewidth}{@{}cccccc@{}}
\toprule
Work                & Model           & AUROC & AUPRC & F1    & Num-Parameters \\ \midrule
\multicolumn{6}{l}{\textit{Algorithmic Baselines}}                             \\
\citet{mesarcik_learning_2022}            & AOFlagger \cite{offringa_aoflagger_2010}       & 0.788 & 0.572 & 0.57  & N/A            \\ \midrule
\multicolumn{6}{l}{\textit{ANN Baselines}}                                     \\
\citet{dutoit_comparison_2024}                & R-Net-5 \cite{sadr_deep_2020}         &       &       & 0.65  & 19.9k          \\
\citet{dutoit_comparison_2024}                & R-Net-7 \cite{sadr_deep_2020}         &       &       & 0.649 & 32.9k          \\
\citet{dutoit_comparison_2024}                & RFI-Net \cite{yang_deep_2020}         &       &       & 0.632 & 809k           \\
\citet{mesarcik_learning_2022}              & U-Net \cite{akeret_radio_2017}           & 0.802 & 0.592 & 0.588 & 292k           \\
\citet{mesarcik_learning_2022}            & AutoEncoder     & 0.862 & 0.622 & 0.511 & 365k                \\
\citet{vanzyl_remove_2024} & RFDL            & 0.989 & 0.748 & 0.675 & 234k               \\
\citet{dutoit_comparison_2024}              & ASPP            &       &       & 0.630 & 292k           \\
\citet{ouyang_hierarchical_2024}              & Swin-UNetR-6M   & 0.971 & 0.678 & 0.630 & 6.5M           \\
\citet{ouyang_hierarchical_2024}              & Swin-UNetR-100M & 0.974 & 0.683 & 0.628 & 102.1M         \\
\citet{ouyang_hierarchical_2024}              & Swin-UNetR-400M & 0.977 & 0.694 & 0.640 & 409.1M         \\ \midrule
\multicolumn{6}{l}{\textit{SNN Baselines}}                                     \\
\citet{pritchard_rfi_2024}           & ANN2SNN         & 0.609 & 0.321 & 0.408 & 692k               \\
\citet{pritchard_spiking_2024}           & BPTT + DN       & 0.346 & 0.604 & 0.474 & 16.4k               \\ \bottomrule
\end{tabularx}
\begin{tablenotes}
    \item This LOFAR dataset is notoriously difficult, and here, while SNN methods lag in detection accuracy, they perform admirably compared to ANN models several orders of magnitude larger in complexity. The LOFAR dataset is also available online \cite{mesarcik_learning_2022}.
\end{tablenotes}
\end{table}
\newpage
\bibliography{sn-bibliography}

@article{pritchard_rfi_2024,
	title = {{RFI} detection with spiking neural networks},
	volume = {41},
	doi = {10.1017/pasa.2024.27},
	journal = {Publications of the Astronomical Society of Australia},
	author = {Pritchard, N.J. and Wicenec, A. and Bennamoun, M. and Dodson, R.},
	year = {2024},
	pages = {e028},
}

@inproceedings{pritchard_supervised_2024,
	title = {Supervised {Radio} {Frequency} {Interference} {Detection} with {SNNs}},
	url = {https://ieeexplore.ieee.org/abstract/document/10766534},
	doi = {10.1109/ICONS62911.2024.00023},
	urldate = {2025-09-16},
	booktitle = {2024 {International} {Conference} on {Neuromorphic} {Systems} ({ICONS})},
	author = {Pritchard, Nicholas J. and Wicenec, Andreas and Bennamoun, Mohammed and Dodson, Richard},
	month = jul,
	year = {2024},
	pages = {102--109},
}

@article{pritchard_polarisation-inclusive_2025,
	title = {Polarization-{Inclusive} {Spiking} {Neural} {Networks} for {Real}-{Time} {RFI} {Detection} in {Modern} {Radio} {Telescopes}},
	volume = {7},
	url = {https://www.ursi.org/Publications/RadioScienceLetters/Volume7/RSL25-0006.pdf},
	doi = {10.46620/25-0006},
	language = {en},
	urldate = {2025-09-22},
	journal = {URSI Radio Science Letters},
	author = {Pritchard, Nicholas J. and Wicenec, Andreas and Dodson, Richard and Bennamoun, Mohammed},
	year = {2025},
}

@article{pritchard_spiking_2024,
	title = {Spiking neural networks for radio frequency interference detection in radio astronomy},
	volume = {8},
	issn = {2399-3650},
	url = {https://www.nature.com/articles/s42005-025-02420-7},
	doi = {10.1038/s42005-025-02420-7},
	language = {en},
	number = {1},
	urldate = {2026-01-09},
	journal = {Communications Physics},
	author = {Pritchard, Nicholas J. and Wicenec, Andreas and Bennamoun, Mohammed and Dodson, Richard},
	month = nov,
	year = {2025},
	pages = {517},
}

@phdthesis{scott_evolving_2015,
    title = {Evolving {Spiking} {Neural} {Networks} for {Spatio}- and {Spectro}- {Temporal} {Data} {Analysis}: {Models}, {Implementations}, {Applications}},
    shorttitle = {Evolving {Spiking} {Neural} {Networks} for {Spatio}- and {Spectro}- {Temporal} {Data} {Analysis}},
    url = {https://hdl.handle.net/10292/10601},
    language = {en},
    urldate = {2024-04-05},
    school = {Auckland University of Technology},
    author = {Scott, Nathan Matthew},
    year = {2015},
}

@article{kasabov_evolving_2016,
    series = {Special {Issue} on "{Neural} {Network} {Learning} in {Big} {Data}"},
    title = {Evolving spatio-temporal data machines based on the {NeuCube} neuromorphic framework: {Design} methodology and selected applications},
    volume = {78},
    issn = {0893-6080},
    shorttitle = {Evolving spatio-temporal data machines based on the {NeuCube} neuromorphic framework},
    url = {https://www.sciencedirect.com/science/article/pii/S0893608015001860},
    doi = {10.1016/j.neunet.2015.09.011},
    urldate = {2024-03-15},
    journal = {Neural Networks},
    author = {Kasabov, Nikola and Scott, Nathan Matthew and Tu, Enmei and Marks, Stefan and Sengupta, Neelava and Capecci, Elisa and Othman, Muhaini and Doborjeh, Maryam Gholami and Murli, Norhanifah and Hartono, Reggio and Espinosa-Ramos, Josafath Israel and Zhou, Lei and Alvi, Fahad Bashir and Wang, Grace and Taylor, Denise and Feigin, Valery and Gulyaev, Sergei and Mahmoud, Mahmoud and Hou, Zeng-Guang and Yang, Jie},
    month = jun,
    year = {2016},
    pages = {1--14},
}

@misc{mesarcik_learning_2022,
    title = {Learning to detect {RFI} in radio astronomy without seeing it},
    url = {https://doi.org/10.5281/zenodo.6724065},
    doi = {10.5281/zenodo.6724065},
    publisher = {Zenodo},
    author = {Mesarcik, Michael and Boonstra, Albert-Jan and van Nieuwpoort, Rob and {Ranguelova}},
    month = jun,
    year = {2022},
}

@article{sadr_deep_2020,
    title = {Deep learning improves identification of {Radio} {Frequency} {Interference}},
    volume = {499},
    issn = {0035-8711},
    doi = {10.1093/mnras/staa2724},
    language = {English},
    number = {1},
    journal = {MONTHLY NOTICES OF THE ROYAL ASTRONOMICAL SOCIETY},
    author = {Sadr, Alireza Vafaei and Bassett, Bruce A. and Oozeer, Nadeem and Fantaye, Yabebal and Finlay, Chris},
    month = nov,
    year = {2020},
    note = {Place: GREAT CLARENDON ST, OXFORD OX2 6DP, ENGLAND
Publisher: OXFORD UNIV PRESS
Type: Article},
    pages = {379--390},
}

@article{akeret_radio_2017,
    title = {Radio frequency interference mitigation using deep convolutional neural networks},
    volume = {18},
    issn = {2213-1337},
    url = {https://www.sciencedirect.com/science/article/pii/S2213133716301056},
    doi = {10.1016/j.ascom.2017.01.002},
    language = {en},
    urldate = {2022-09-20},
    journal = {Astronomy and Computing},
    author = {Akeret, J. and Chang, C. and Lucchi, A. and Refregier, A.},
    month = jan,
    year = {2017},
    pages = {35--39},
}

@article{vanzyl_remove_2024,
    title = {Remove {First} {Detect} {Later}: a counter-intuitive approach for detecting radio frequency interference in radio sky imagery},
    volume = {530},
    issn = {0035-8711},
    shorttitle = {Remove {First} {Detect} {Later}},
    url = {https://doi.org/10.1093/mnras/stae979},
    doi = {10.1093/mnras/stae979},
    number = {2},
    urldate = {2024-10-23},
    journal = {Monthly Notices of the Royal Astronomical Society},
    author = {van Zyl, Daniel J and Grobler, Trienko L},
    month = may,
    year = {2024},
    pages = {1907--1920},
}

@article{dutoit_comparison_2024,
    title = {A comparison framework for deep learning {RFI} detection algorithms},
    volume = {530},
    issn = {0035-8711},
    url = {https://doi.org/10.1093/mnras/stae892},
    doi = {10.1093/mnras/stae892},
    number = {1},
    urldate = {2024-04-29},
    journal = {Monthly Notices of the Royal Astronomical Society},
    author = {Du Toit, Charl D and Grobler, Trienko L and Ludick, Danie J},
    month = may,
    year = {2024},
    pages = {613--629},
}

@inproceedings{pritchard_advancing_2025,
	title = {Advancing {RFI}-{Detection} in {Radio} {Astronomy} with {Liquid} {State} {Machines}},
	url = {https://ieeexplore.ieee.org/document/11229299},
	doi = {10.1109/IJCNN64981.2025.11229299},
	urldate = {2025-11-17},
	booktitle = {2025 {International} {Joint} {Conference} on {Neural} {Networks} ({IJCNN})},
	author = {Pritchard, Nicholas J. and Wicenec, Andreas and Bennamoun, Mohammed and Dodson, Richard},
	month = jun,
	year = {2025},
	note = {ISSN: 2161-4407},
	pages = {1--7},
}

@inproceedings{ouyang_hierarchical_2024,
    address = {Bariloche, Argentina},
    title = {Hierarchical vision transformers for {RFI} mitigation in radio astronomy},
    language = {en},
    author = {Ouyang, Xiaowei and Dreuning, Henk and Mesarcik, Michael and van Nieuwpoort, Rob V},
    month = oct,
    year = {2024},
}

@article{offringa_aoflagger_2010,
    title = {{AOFlagger}: {RFI} {Software}},
    journal = {Astrophysics Source Code Library},
    author = {Offringa, AR},
    year = {2010},
    pages = {ascl--1010},
}

@article{vermij_challenges_2015,
    title = {Challenges in exascale radio astronomy: {Can} the {SKA} ride the technology wave?},
    volume = {29},
    issn = {1094-3420},
    shorttitle = {Challenges in exascale radio astronomy},
    url = {https://doi.org/10.1177/1094342014549059},
    doi = {10.1177/1094342014549059},
    language = {en},
    number = {1},
    urldate = {2024-10-28},
    journal = {The International Journal of High Performance Computing Applications},
    author = {Vermij, Erik and Fiorin, Leandro and Jongerius, Rik and Hagleitner, Christoph and Bertels, Koen},
    month = feb,
    year = {2015},
    note = {Publisher: SAGE Publications Ltd STM},
    pages = {37--50},
}

@article{schuman_opportunities_2022,
    title = {Opportunities for neuromorphic computing algorithms and applications},
    volume = {2},
    copyright = {2022 Springer Nature America, Inc.},
    issn = {2662-8457},
    url = {https://www.nature.com/articles/s43588-021-00184-y},
    doi = {10.1038/s43588-021-00184-y},
    language = {en},
    number = {1},
    urldate = {2023-11-24},
    journal = {Nature Computational Science},
    author = {Schuman, Catherine D. and Kulkarni, Shruti R. and Parsa, Maryam and Mitchell, J. Parker and Date, Prasanna and Kay, Bill},
    month = jan,
    year = {2022},
    note = {Number: 1
Publisher: Nature Publishing Group},
    pages = {10--19},
}

@article{karamimanesh_spiking_2025,
    title = {Spiking neural networks on {FPGA}: {A} survey of methodologies and recent advancements},
    volume = {186},
    issn = {0893-6080},
    shorttitle = {Spiking neural networks on {FPGA}},
    url = {https://www.sciencedirect.com/science/article/pii/S0893608025001352},
    doi = {10.1016/j.neunet.2025.107256},
    urldate = {2025-02-19},
    journal = {Neural Networks},
    author = {Karamimanesh, Mehrzad and Abiri, Ebrahim and Shahsavari, Mahyar and Hassanli, Kourosh and van Schaik, André and Eshraghian, Jason},
    month = jun,
    year = {2025},
    pages = {107256},
}

@article{frenkel_bottom-up_2023,
    title = {Bottom-{Up} and {Top}-{Down} {Approaches} for the {Design} of {Neuromorphic} {Processing} {Systems}: {Tradeoffs} and {Synergies} {Between} {Natural} and {Artificial} {Intelligence}},
    volume = {111},
    issn = {1558-2256},
    shorttitle = {Bottom-{Up} and {Top}-{Down} {Approaches} for the {Design} of {Neuromorphic} {Processing} {Systems}},
    url = {https://ieeexplore.ieee.org/document/10144567},
    doi = {10.1109/JPROC.2023.3273520},
    number = {6},
    urldate = {2024-03-13},
    journal = {Proceedings of the IEEE},
    author = {Frenkel, Charlotte and Bol, David and Indiveri, Giacomo},
    month = jun,
    year = {2023},
    note = {Conference Name: Proceedings of the IEEE},
    pages = {623--652},
}

@misc{pritchard_neuromorphic_2025,
	title = {Neuromorphic {Astronomy}: {An} {End}-to-{End} {SNN} {Pipeline} for {RFI} {Detection} {Hardware}},
	shorttitle = {Neuromorphic {Astronomy}},
	url = {http://arxiv.org/abs/2511.16060},
	doi = {10.48550/arXiv.2511.16060},
	urldate = {2025-11-21},
	publisher = {arXiv},
	author = {Pritchard, Nicholas J. and Wicenec, Andreas and Dodson, Richard and Bennamoun, Mohammed and Muir, Dylan R.},
	month = nov,
	year = {2025},
	note = {arXiv:2511.16060 [cs]},
}

@article{yang_deep_2020,
	title = {Deep residual detection of radio frequency interference for {FAST}},
	volume = {492},
	issn = {0035-8711},
	url = {https://doi.org/10.1093/mnras/stz3521},
	doi = {10.1093/mnras/stz3521},
	number = {1},
	urldate = {2023-08-16},
	journal = {Monthly Notices of the Royal Astronomical Society},
	author = {Yang, Zhicheng and Yu, Ce and Xiao, Jian and Zhang, Bo},
	month = feb,
	year = {2020},
	pages = {1421--1431},
}

@article{morrison_mwax_2023,
    title = {{MWAX}: {A} {New} {Correlator} for the {Murchison} {Widefield} {Array}},
    volume = {40},
    issn = {1323-3580, 1448-6083},
    shorttitle = {{MWAX}},
    url = {http://arxiv.org/abs/2303.11557},
    doi = {10.1017/pasa.2023.15},
    urldate = {2025-09-17},
    journal = {Publications of the Astronomical Society of Australia},
    author = {Morrison, I. S. and Crosse, B. and Sleap, G. and Wayth, R. B. and Williams, A. and Johnston-Hollitt, M. and Jones, J. and Tingay, S. J. and Walker, M. and Williams, L.},
    year = {2023},
    note = {arXiv:2303.11557 [astro-ph]},
    pages = {e019},
}

@article{prabu_digital-receiver_2015,
    title = {A digital-receiver for the {MurchisonWidefield} {Array}},
    volume = {39},
    issn = {1572-9508},
    url = {https://doi.org/10.1007/s10686-015-9444-3},
    doi = {10.1007/s10686-015-9444-3},
    language = {en},
    number = {1},
    urldate = {2025-09-24},
    journal = {Experimental Astronomy},
    author = {Prabu, Thiagaraj and Srivani, K. S. and Roshi, D. Anish and Kamini, P. A. and Madhavi, S. and Emrich, David and Crosse, Brian and Williams, Andrew J. and Waterson, Mark and Deshpande, Avinash A. and Shankar, N. Udaya and Subrahmanyan, Ravi and Briggs, Frank H. and Goeke, Robert F. and Tingay, Steven J. and Johnston-Hollitt, Melanie and R, Gopalakrishna M. and Morgan, Edward H. and Pathikulangara, Joseph and Bunton, John D. and Hampson, Grant and Williams, Christopher and Ord, Stephen M. and Wayth, Randall B. and Kumar, Deepak and Morales, Miguel F. and deSouza, Ludi and Kratzenberg, Eric and Pallot, D. and McWhirter, Russell and Hazelton, Bryna J. and Arcus, Wayne and Barnes, David G. and Bernardi, Gianni and Booler, T. and Bowman, Judd D. and Cappallo, Roger J. and Corey, Brian E. and Greenhill, Lincoln J. and Herne, David and Hewitt, Jacqueline N. and Kaplan, David L. and Kasper, Justin C. and Kincaid, Barton B. and Koenig, Ronald and Lonsdale, Colin J. and Lynch, Mervyn J. and Mitchell, Daniel A. and Oberoi, Divya and Remillard, Ronald A. and Rogers, Alan E. and Salah, Joseph E. and Sault, Robert J. and Stevens, Jamie B. and Tremblay, S. and Webster, Rachel L. and Whitney, Alan R. and Wyithe, Stuart B.},
    month = mar,
    year = {2015},
    pages = {73--93},
}

@article{girish_progression_2023,
    title = {Progression of digital-receiver architecture: {From} {MWA} to {SKA1}-{Low}, and beyond},
    volume = {44},
    issn = {0973-7758},
    shorttitle = {Progression of digital-receiver architecture},
    url = {https://doi.org/10.1007/s12036-023-09921-3},
    doi = {10.1007/s12036-023-09921-3},
    language = {en},
    number = {1},
    urldate = {2025-09-24},
    journal = {Journal of Astrophysics and Astronomy},
    author = {Girish, B. S. and Reddy, S. Harshavardhan and Sethi, Shiv and Srivani, K. S. and Abhishek, R. and Ajithkumar, B. and Bhattramakki, Sahana and Buch, Kaushal and Chaudhuri, Sandeep and Gupta, Yashwant and Kamini, P. A. and Kudale, Sanjay and Madhavi, S. and Muley, Mekhala and Prabu, T. and Raghunathan, Agaram and Shelton, G. J.},
    month = mar,
    year = {2023},
    pages = {28},
}

@inproceedings{nazeer_language_2024,
    title = {Language {Modeling} on a {SpiNNaker2} {Neuromorphic} {Chip}},
    url = {https://ieeexplore.ieee.org/abstract/document/10595870},
    doi = {10.1109/AICAS59952.2024.10595870},
    urldate = {2025-06-13},
    booktitle = {2024 {IEEE} 6th {International} {Conference} on {AI} {Circuits} and {Systems} ({AICAS})},
    author = {Nazeer, Khaleelulla Khan and Schöne, Mark and Mukherji, Rishav and Vogginger, Bernhard and Mayr, Christian and Kappel, David and Subramoney, Anand},
    month = apr,
    year = {2024},
    note = {ISSN: 2834-9857},
    pages = {492--496},
}

@article{wayth_phase_2018,
    title = {The {Phase} {II} {Murchison} {Widefield} {Array}: {Design} {Overview}},
    volume = {35},
    issn = {1323-3580, 1448-6083},
    shorttitle = {The {Phase} {II} {Murchison} {Widefield} {Array}},
    url = {http://arxiv.org/abs/1809.06466},
    doi = {10.1017/pasa.2018.37},
    urldate = {2025-09-24},
    journal = {Publications of the Astronomical Society of Australia},
    author = {Wayth, Randall B. and Tingay, Steven J. and Trott, Cathryn M. and Emrich, David and Johnston-Hollitt, Melanie and McKinley, Ben and Gaensler, B. M. and Beardsley, A. P. and Booler, T. and Crosse, B. and Franzen, T. M. O. and Horsley, L. and Kaplan, D. L. and Kenney, D. and Morales, M. F. and Pallot, D. and Sleap, G. and Steele, K. and Walker, M. and Williams, A. and Wu, C. and Cairns, Iver H. and Filipovic, M. D. and Johnston, S. and Murphy, T. and Quinn, P. and Staveley-Smith, L. and Webster, R. and Wyithe, J. S. B.},
    year = {2018},
    note = {arXiv:1809.06466 [astro-ph]},
    pages = {e033},
}

@article{noauthor_nvidia_nodate,
    title = {{NVIDIA} {A40} datasheet},
    language = {en},
}

@article{hotan_australian_2021,
    title = {Australian square kilometre array pathfinder: {I}. system description},
    volume = {38},
    issn = {1323-3580, 1448-6083},
    shorttitle = {Australian square kilometre array pathfinder},
    url = {https://www.cambridge.org/core/journals/publications-of-the-astronomical-society-of-australia/article/australian-square-kilometre-array-pathfinder-i-system-description/E1A1DC57B4850811F5624AD3E97C9C9E},
    doi = {10.1017/pasa.2021.1},
    language = {en},
    urldate = {2025-09-26},
    journal = {Publications of the Astronomical Society of Australia},
    author = {Hotan, A. W. and Bunton, J. D. and Chippendale, A. P. and Whiting, M. and Tuthill, J. and Moss, V. A. and McConnell, D. and Amy, S. W. and Huynh, M. T. and Allison, J. R. and Anderson, C. S. and Bannister, K. W. and Bastholm, E. and Beresford, R. and Bock, D. C.-J. and Bolton, R. and Chapman, J. M. and Chow, K. and Collier, J. D. and Cooray, F. R. and Cornwell, T. J. and Diamond, P. J. and Edwards, P. G. and Feain, I. J. and Franzen, T. M. O. and George, D. and Gupta, N. and Hampson, G. A. and Harvey-Smith, L. and Hayman, D. B. and Heywood, I. and Jacka, C. and Jackson, C. A. and Jackson, S. and Jeganathan, K. and Johnston, S. and Kesteven, M. and Kleiner, D. and Koribalski, B. S. and Lee-Waddell, K. and Lenc, E. and Lensson, E. S. and Mackay, S. and Mahony, E. K. and McClure-Griffiths, N. M. and McConigley, R. and Mirtschin, P. and Ng, A. K. and Norris, R. P. and Pearce, S. E. and Phillips, C. and Pilawa, M. A. and Raja, W. and Reynolds, J. E. and Roberts, P. and Roxby, D. N. and Sadler, E. M. and Shields, M. and Schinckel, A. E. T. and Serra, P. and Shaw, R. D. and Sweetnam, T. and Troup, E. R. and Tzioumis, A. and Voronkov, M. A. and Westmeier, T.},
    month = jan,
    year = {2021},
    pages = {e009},
}

@inproceedings{brown_design_2014,
    title = {Design and implementation of the 2nd {Generation} {ASKAP} {Digital} {Receiver} {System}},
    url = {https://ieeexplore.ieee.org/document/6903860},
    doi = {10.1109/ICEAA.2014.6903860},
    urldate = {2025-10-02},
    booktitle = {2014 {International} {Conference} on {Electromagnetics} in {Advanced} {Applications} ({ICEAA})},
    author = {Brown, A. J. and Hampson, G.A. and Roberts, P. and Beresford, R. and Bunton, J.D. and Cheng, W. and Chekkala, R. and Kiraly, D. and Neuhold, S. and Jeganathan, K.},
    month = aug,
    year = {2014},
    pages = {268--271},
}

@inproceedings{hampson_askap_2014,
    address = {Beijing, China},
    title = {{ASKAP} {Redback}-3 \&\#x2014; {An} agile digital signal processing platform},
    isbn = {978-1-4673-5225-3},
    url = {http://ieeexplore.ieee.org/document/6930062/},
    doi = {10.1109/URSIGASS.2014.6930062},
    language = {en},
    urldate = {2025-10-02},
    booktitle = {2014 {XXXIth} {URSI} {General} {Assembly} and {Scientific} {Symposium} ({URSI} {GASS})},
    publisher = {IEEE},
    author = {Hampson, G. A. and Brown, A. and Bunton, J. D. and Neuhold, S. and Chekkala, R. and Bateman, T. and Tuthill, J.},
    month = aug,
    year = {2014},
    pages = {1--4},
}

@article{wang_craft_2025,
    title = {The {CRAFT} {Coherent} ({CRACO}) upgrade {I}: {System} {Description} and {Results} of the 110-ms {Radio} {Transient} {Pilot} {Survey}},
    volume = {42},
    issn = {1323-3580, 1448-6083},
    shorttitle = {The {CRAFT} {Coherent} ({CRACO}) upgrade {I}},
    url = {http://arxiv.org/abs/2409.10316},
    doi = {10.1017/pasa.2024.107},
    language = {en},
    urldate = {2025-10-02},
    journal = {Publications of the Astronomical Society of Australia},
    author = {Wang, Z. and Bannister, K. W. and Gupta, V. and Deng, X. and Pilawa, M. and Tuthill, J. and Bunton, J. D. and Flynn, C. and Glowacki, M. and Jaini, A. and Lee, Y. W. J. and Lenc, E. and Lucero, J. and Paek, A. and Radhakrishnan, R. and Thyagarajan, N. and Uttarkar, P. and Wang, Y. and Bhat, N. D. R. and James, C. W. and Moss, V. A. and Murphy, Tara and Reynolds, J. E. and Shannon, R. M. and Spitler, L. G. and Tzioumis, A. and Caleb, M. and Deller, A. T. and Gordon, A. C. and Marnoch, L. and Ryder, S. D. and Simha, S. and Anderson, C. S. and Ball, L. and Brodrick, D. and Cooray, F. R. and Gupta, N. and Hayman, D. B. and Ng, A. and Pearce, S. E. and Phillips, C. and Voronkov, M. A. and Westmeier, T.},
    year = {2025},
    note = {arXiv:2409.10316 [astro-ph]},
    pages = {e005},
}

@inproceedings{whiting_high-performance_2018,
    address = {Gran Canaria},
    title = {High-{Performance} {Pipeline} {Processing} for {ASKAP}},
    isbn = {978-90-825987-3-5},
    url = {https://ieeexplore.ieee.org/document/8471521/},
    doi = {10.23919/URSI-AT-RASC.2018.8471521},
    language = {en},
    urldate = {2025-10-03},
    booktitle = {2018 2nd {URSI} {Atlantic} {Radio} {Science} {Meeting} ({AT}-{RASC})},
    publisher = {IEEE},
    author = {Whiting, M. and Ord, S. M. and Mitchell, D. and Voronkov, M. and Guzman, J. C.},
    month = may,
    year = {2018},
    pages = {1--1},
}

@article{guzman_are_nodate,
    title = {Are we {There} {Yet}? {Experiences} from {Developing} and {Commissioning} the {High} {Performance} {Computing} ({HPC}) {System} for the {ASKAP} {Telescope}},
    language = {en},
    author = {Guzman, Juan C and Bastholm, Eric and Raja, Wasim and Whiting, Matthew and Mitchell, Daniel and Voronkov, Max and Ord, Stephen},
}

@article{van_haarlem_lofar_2013,
    title = {{LOFAR}: {The} {LOw}-{Frequency} {ARray}},
    volume = {556},
    issn = {0004-6361, 1432-0746},
    shorttitle = {{LOFAR}},
    url = {http://www.aanda.org/10.1051/0004-6361/201220873},
    doi = {10.1051/0004-6361/201220873},
    language = {en},
    urldate = {2025-09-17},
    journal = {Astronomy \& Astrophysics},
    author = {Van Haarlem, M. P. and Wise, M. W. and Gunst, A. W. and Heald, G. and McKean, J. P. and Hessels, J. W. T. and De Bruyn, A. G. and Nijboer, R. and Swinbank, J. and Fallows, R. and Brentjens, M. and Nelles, A. and Beck, R. and Falcke, H. and Fender, R. and Hörandel, J. and Koopmans, L. V. E. and Mann, G. and Miley, G. and Röttgering, H. and Stappers, B. W. and Wijers, R. A. M. J. and Zaroubi, S. and Van Den Akker, M. and Alexov, A. and Anderson, J. and Anderson, K. and Van Ardenne, A. and Arts, M. and Asgekar, A. and Avruch, I. M. and Batejat, F. and Bähren, L. and Bell, M. E. and Bell, M. R. and Van Bemmel, I. and Bennema, P. and Bentum, M. J. and Bernardi, G. and Best, P. and Bîrzan, L. and Bonafede, A. and Boonstra, A.-J. and Braun, R. and Bregman, J. and Breitling, F. and Van De Brink, R. H. and Broderick, J. and Broekema, P. C. and Brouw, W. N. and Brüggen, M. and Butcher, H. R. and Van Cappellen, W. and Ciardi, B. and Coenen, T. and Conway, J. and Coolen, A. and Corstanje, A. and Damstra, S. and Davies, O. and Deller, A. T. and Dettmar, R.-J. and Van Diepen, G. and Dijkstra, K. and Donker, P. and Doorduin, A. and Dromer, J. and Drost, M. and Van Duin, A. and Eislöffel, J. and Van Enst, J. and Ferrari, C. and Frieswijk, W. and Gankema, H. and Garrett, M. A. and De Gasperin, F. and Gerbers, M. and De Geus, E. and Grießmeier, J.-M. and Grit, T. and Gruppen, P. and Hamaker, J. P. and Hassall, T. and Hoeft, M. and Holties, H. A. and Horneffer, A. and Van Der Horst, A. and Van Houwelingen, A. and Huijgen, A. and Iacobelli, M. and Intema, H. and Jackson, N. and Jelic, V. and De Jong, A. and Juette, E. and Kant, D. and Karastergiou, A. and Koers, A. and Kollen, H. and Kondratiev, V. I. and Kooistra, E. and Koopman, Y. and Koster, A. and Kuniyoshi, M. and Kramer, M. and Kuper, G. and Lambropoulos, P. and Law, C. and Van Leeuwen, J. and Lemaitre, J. and Loose, M. and Maat, P. and Macario, G. and Markoff, S. and Masters, J. and McFadden, R. A. and McKay-Bukowski, D. and Meijering, H. and Meulman, H. and Mevius, M. and Middelberg, E. and Millenaar, R. and Miller-Jones, J. C. A. and Mohan, R. N. and Mol, J. D. and Morawietz, J. and Morganti, R. and Mulcahy, D. D. and Mulder, E. and Munk, H. and Nieuwenhuis, L. and Van Nieuwpoort, R. and Noordam, J. E. and Norden, M. and Noutsos, A. and Offringa, A. R. and Olofsson, H. and Omar, A. and Orrú, E. and Overeem, R. and Paas, H. and Pandey-Pommier, M. and Pandey, V. N. and Pizzo, R. and Polatidis, A. and Rafferty, D. and Rawlings, S. and Reich, W. and De Reijer, J.-P. and Reitsma, J. and Renting, G. A. and Riemers, P. and Rol, E. and Romein, J. W. and Roosjen, J. and Ruiter, M. and Scaife, A. and Van Der Schaaf, K. and Scheers, B. and Schellart, P. and Schoenmakers, A. and Schoonderbeek, G. and Serylak, M. and Shulevski, A. and Sluman, J. and Smirnov, O. and Sobey, C. and Spreeuw, H. and Steinmetz, M. and Sterks, C. G. M. and Stiepel, H.-J. and Stuurwold, K. and Tagger, M. and Tang, Y. and Tasse, C. and Thomas, I. and Thoudam, S. and Toribio, M. C. and Van Der Tol, B. and Usov, O. and Van Veelen, M. and Van Der Veen, A.-J. and Ter Veen, S. and Verbiest, J. P. W. and Vermeulen, R. and Vermaas, N. and Vocks, C. and Vogt, C. and De Vos, M. and Van Der Wal, E. and Van Weeren, R. and Weggemans, H. and Weltevrede, P. and White, S. and Wijnholds, S. J. and Wilhelmsson, T. and Wucknitz, O. and Yatawatta, S. and Zarka, P. and Zensus, A. and Van Zwieten, J.},
    month = aug,
    year = {2013},
    pages = {A2},
}

@article{schoonderbeek_uniboard2_2019,
    title = {{UniBoard2}, {A} {Generic} {Scalable} {High}-{Performance} {Computing} {Platform} for {Radio} {Astronomy}},
    volume = {08},
    issn = {2251-1717},
    url = {https://www.worldscientific.com/doi/full/10.1142/S225117171950003X},
    doi = {10.1142/S225117171950003X},
    number = {02},
    urldate = {2025-10-02},
    journal = {Journal of Astronomical Instrumentation},
    author = {Schoonderbeek, G. W. and Szomoru, A. and Gunst, A. W. and Hiemstra, L. and Hargreaves, J.},
    month = jun,
    year = {2019},
    note = {Publisher: World Scientific Publishing Co.},
    pages = {1950003},
}

@article{broekema_cobalt_2018,
    title = {Cobalt: {A} {GPU}-based correlator and beamformer for {LOFAR}},
    volume = {23},
    issn = {2213-1337},
    shorttitle = {Cobalt},
    url = {https://www.sciencedirect.com/science/article/pii/S2213133717301439},
    doi = {10.1016/j.ascom.2018.04.006},
    urldate = {2025-10-02},
    journal = {Astronomy and Computing},
    author = {Broekema, P. Chris and Mol, J. Jan David and Nijboer, R. and van Amesfoort, A. S. and Brentjens, M. A. and Loose, G. Marcel and Klijn, W. F. A. and Romein, J. W.},
    month = apr,
    year = {2018},
    pages = {180--192},
}

@article{juerges_lofar20_2022,
    title = {{LOFAR2}.0: {Station} {Control} {Upgrade}},
    volume = {ICALEPCS2021},
    copyright = {Creative Commons Attribution 3.0 Unported},
    issn = {2226-0358},
    shorttitle = {{LOFAR2}.0},
    url = {https://jacow.org/icalepcs2021/doi/JACoW-ICALEPCS2021-MOAR03.html},
    doi = {10.18429/JACOW-ICALEPCS2021-MOAR03},
    language = {en},
    urldate = {2025-10-03},
    journal = {Proceedings of the 18th International Conference on Accelerator and Large Experimental Physics Control Systems},
    author = {Juerges, Thomas and Mol, Jan David and Snijder, Thijs},
    editor = {Kazuro (Ed.), Furukawa and Yingbing (Ed.), Yan and Yongbin (Ed.), Leng and Zhichu (Ed.), Chen and R.W. (Ed.), Volker, Schaa},
    year = {2022},
    note = {Artwork Size: 6 pages, 0.469 MB
ISBN: 9783954502219
Medium: PDF
Publisher: JACoW Publishing, Geneva, Switzerland},
    pages = {6 pages, 0.469 MB},
}

@inproceedings{chaudhari_station_2024,
    address = {Bhimtal, India},
    title = {Station {Signal} {Processing} {System} for the {SKA}-{Low}: status, plans and {India}'s role},
    isbn = {978-94-63968-13-3},
    shorttitle = {Station {Signal} {Processing} {System} for the {SKA}-{Low}},
    url = {https://www.ursi.org/proceedings/RCRS/2024/RCRS2024\_0997.pdf},
    doi = {10.46620/URSI_RSRC24/0997NNP3287},
    language = {en},
    urldate = {2025-10-02},
    booktitle = {Proceedings of the 6th {URSI} {Regional} {Conference} on {Radio} {Science} – {RCRS} 2024},
    publisher = {URSI – International Union of Radio Science},
    author = {Chaudhari, Sandeep and Gupta, Yashwant and T, Prabu},
    year = {2024},
}

@article{benthem_aperture_2021,
    title = {The {Aperture} {Array} {Verification} {System} 1: {System} overview and early commissioning results},
    volume = {655},
    copyright = {© ESO 2021},
    issn = {0004-6361, 1432-0746},
    shorttitle = {The {Aperture} {Array} {Verification} {System} 1},
    url = {https://www.aanda.org/articles/aa/abs/2021/11/aa40086-20/aa40086-20.html},
    doi = {10.1051/0004-6361/202040086},
    language = {en},
    urldate = {2025-10-02},
    journal = {Astronomy \& Astrophysics},
    author = {Benthem, P. and Wayth, R. and Acedo, E. de Lera and Adami, K. Zarb and Alderighi, M. and Belli, C. and Bolli, P. and Booler, T. and Borg, J. and Broderick, J. W. and Chiarucci, S. and Chiello, R. and Ciani, L. and Comoretto, G. and Crosse, B. and Davidson, D. and DeMarco, A. and Emrich, D. and Es, A. van and Fierro, D. and Faulkner, A. and Gerbers, M. and Razavi-Ghods, N. and Hall, P. and Horsley, L. and Juswardy, B. and Kenney, D. and Steele, K. and Magro, A. and Mattana, A. and McKinley, B. and Monari, J. and Naldi, G. and Nanni, J. and Ninni, P. Di and Paonessa, F. and Perini, F. and Poloni, M. and Pupillo, G. and Rusticelli, S. and Schiaffino, M. and Schillirò, F. and Schnetler, H. and Singuaroli, R. and Sokolowski, M. and Sutinjo, A. and Tartarini, G. and Ung, D. and Vaate, J. G. Bij de and Virone, G. and Walker, M. and Waterson, M. and Wijnholds, S. J. and Williams, A.},
    month = nov,
    year = {2021},
    note = {Publisher: EDP Sciences},
    pages = {A5},
}

@article{hampson_square_2022,
    title = {Square {Kilometre} {Array} {Low} {Atomic} commercial off-the-shelf correlator and beamformer},
    volume = {8},
    issn = {2329-4124},
    url = {https://www.spiedigitallibrary.org/journals/Journal-of-Astronomical-Telescopes-Instruments-and-Systems/volume-8/issue-01/011018/Square-Kilometre-Array-Low-Atomic-commercial-off-the-shelf-correlator/10.1117/1.JATIS.8.1.011018.full},
    doi = {10.1117/1.JATIS.8.1.011018},
    language = {en},
    number = {01},
    urldate = {2025-10-02},
    journal = {Journal of Astronomical Telescopes, Instruments, and Systems},
    author = {Hampson, Grant A. and Bunton, John D. and Humphrey, David and Bengston, Keith J. and Jourjon, Guillaume and Bolin, Andrew B. and Chen, Yuqing and Troup, Euan R. and Babich, Giles C. and Van Aardt, Jason C.},
    month = feb,
    year = {2022},
}

@misc{williamson_optimising_2024,
    title = {Optimising the {Processing} and {Storage} of {Radio} {Astronomy} {Data}},
    url = {http://arxiv.org/abs/2410.02285},
    doi = {10.48550/arXiv.2410.02285},
    urldate = {2025-09-17},
    publisher = {arXiv},
    author = {Williamson, Alexander and Elahi, Pascal J. and Dodson, Richard and Rhee, Jonghwan and Gong, Qian},
    month = oct,
    year = {2024},
    note = {arXiv:2410.02285 [astro-ph]
version: 1},
}

@inproceedings{pleasance_talon_2021,
    title = {{TALON} {Demonstration} {Correlator} {Architecture} for {Early} {SKA} {Array} {Assemblies}},
    url = {https://ieeexplore.ieee.org/abstract/document/9560125},
    doi = {10.23919/URSIGASS51995.2021.9560125},
    urldate = {2025-10-02},
    booktitle = {2021 {XXXIVth} {General} {Assembly} and {Scientific} {Symposium} of the {International} {Union} of {Radio} {Science} ({URSI} {GASS})},
    author = {Pleasance, Mike and Carlson, Brent and Vrcic, Sonja and Gunaratne, Thushara},
    month = aug,
    year = {2021},
    note = {ISSN: 2642-4339},
    pages = {1--4},
}

@inproceedings{roy_overview_2022,
	title = {Overview of the {Single} {Pixel} {Feed} {Receiver} {System} of {Square} {Kilometer} {Array} {MID} {Telescope}},
	url = {https://ieeexplore.ieee.org/abstract/document/9814339},
	doi = {10.23919/AT-AP-RASC54737.2022.9814339},
	urldate = {2025-10-03},
	booktitle = {2022 3rd {URSI} {Atlantic} and {Asia} {Pacific} {Radio} {Science} {Meeting} ({AT}-{AP}-{RASC})},
	author = {Roy, Jayashree and Pleasance, Michael and Harrison, Stephen and Wolfgang, Andreas and Caputa, Kris},
	month = may,
	year = {2022},
	pages = {1--4},
}

@inproceedings{hall_innovative_2016,
    author = {Gie Han Tan and Robert Lehmensiek and Bhushan Billade and Krzysztof Caputa and St{\'e}phane Gauffre and Isak P. Theron and Miroslav Pantaleev and Zoran Ljusic and Benjamin Quertier and Adriaan Peens-Hough},
    title = {{An innovative, highly sensitive receiver system for the Square Kilometre Array Mid Radio Telescope}},
    volume = {9906},
    booktitle = {Ground-based and Airborne Telescopes VI},
    editor = {Helen J. Hall and Roberto Gilmozzi and Heather K. Marshall},
    organization = {International Society for Optics and Photonics},
    publisher = {SPIE},
    pages = {990660},
    year = {2016},
    doi = {10.1117/12.2230897},
    URL = {https://doi.org/10.1117/12.2230897}
}

@article{swinbank_lofar_2015,
    title = {The {LOFAR} {Transients} {Pipeline}},
    volume = {11},
    issn = {2213-1337},
    url = {https://www.sciencedirect.com/science/article/pii/S2213133715000207},
    doi = {10.1016/j.ascom.2015.03.002},
    urldate = {2025-10-03},
    journal = {Astronomy and Computing},
    author = {Swinbank, John D. and Staley, Tim D. and Molenaar, Gijs J. and Rol, Evert and Rowlinson, Antonia and Scheers, Bart and Spreeuw, Hanno and Bell, Martin E. and Broderick, Jess W. and Carbone, Dario and Garsden, Hugh and van der Horst, Alexander J. and Law, Casey J. and Wise, Michael and Breton, Rene P. and Cendes, Yvette and Corbel, Stéphane and Eislöffel, Jochen and Falcke, Heino and Fender, Rob and Grießmeier, Jean-Mathias and Hessels, Jason W. T. and Stappers, Benjamin W. and Stewart, Adam J. and Wijers, Ralph A. M. J. and Wijnands, Rudy and Zarka, Philippe},
    month = jun,
    year = {2015},
    pages = {25--48},
}

@article{sokolowski_southern-hemisphere_2021,
    title = {A {Southern}-{Hemisphere} all-sky radio transient monitor for {SKA}-{Low} prototype stations},
    volume = {38},
    issn = {1323-3580, 1448-6083},
    url = {https://www.cambridge.org/core/journals/publications-of-the-astronomical-society-of-australia/article/southernhemisphere-allsky-radio-transient-monitor-for-skalow-prototype-stations/E921D04E346E3C745C0E4D9D59096A3C},
    doi = {10.1017/pasa.2021.16},
    language = {en},
    urldate = {2025-10-03},
    journal = {Publications of the Astronomical Society of Australia},
    author = {Sokolowski, M. and Wayth, R. B. and Bhat, N. D. R. and Price, D. and Broderick, J. W. and Bernardi, G. and Bolli, P. and Chiello, R. and Comoretto, G. and Crosse, B. and Davidson, D. B. and Macario, G. and Magro, A. and Mattana, A. and Minchin, D. and McPhail, A. and Monari, J. and Perini, F. and Pupillo, G. and Sleap, G. and Tingay, S. and Ung, D. and Williams, A.},
    month = jan,
    year = {2021},
    pages = {e023},
}

@book{thompson_interferometry_2017,
    address = {Cham},
    edition = {3rd ed. 2017.},
    series = {Astronomy and {Astrophysics} {Library}},
    title = {Interferometry and {Synthesis} in {Radio} {Astronomy}},
    isbn = {3-319-44431-X},
    language = {eng},
    publisher = {Springer Nature},
    author = {Thompson, A. Richard. and Moran, James M. and Swenson Jr., George W.},
    year = {2017},
    note = {Publication Title: Interferometry and Synthesis in Radio Astronomy},
}

@article{aafreen_high-performance_2023,
    title = {High-performance computing for {SKA} transient search: {Use} of {FPGA}-based accelerators},
    volume = {44},
    issn = {0973-7758},
    shorttitle = {High-performance computing for {SKA} transient search},
    url = {https://doi.org/10.1007/s12036-022-09896-7},
    doi = {10.1007/s12036-022-09896-7},
    language = {en},
    number = {1},
    urldate = {2025-10-03},
    journal = {Journal of Astrophysics and Astronomy},
    author = {Aafreen, R. and Abhishek, R. and Ajithkumar, B. and Vaidyanathan, Arunkumar M. and Barve, Indrajit V. and Bhattramakki, Sahana and Bhat, Shashank and Girish, B. S. and Ghalame, Atul and Gupta, Y. and Hayatnagarkar, Harshal G. and Kamini, P. A. and Karastergiou, A. and Levin, L. and Madhavi, S. and Mekhala, M. and Mickaliger, M. and Mugundhan, V. and Naidu, Arun and Oppermann, J. and Pandian, B. Arul and Patra, N. and Raghunathan, A. and Roy, Jayanta and Sethi, Shiv and Shaw, B. and Sherwin, K. and Sinnen, O. and Sinha, S. K. and Srivani, K. S. and Stappers, B. and Subrahmanya, C. R. and Prabu, Thiagaraj and Vinutha, C. and Wadadekar, Y. G. and Wang, Haomiao and Williams, C.},
    month = feb,
    year = {2023},
    pages = {11},
}

@article{davies_intel_nodate,
    title = {Intel {Labs}’ new {Loihi} 2 research chip outperforms its predecessor by up to 10x and comes with an open-source, community-driven neuromorphic computing framework},
    url = {https://download.intel.com/newsroom/2021/new-technologies/neuromorphic-computing-loihi-2-brief.pdf},
    language = {en},
    author = {Davies, Mike},
}

@article{swart_highlights_2022,
    title = {Highlights of the {SKA1}-{Mid} telescope architecture},
    volume = {8},
    issn = {2329-4124, 2329-4221},
    url = {https://www.spiedigitallibrary.org/journals/Journal-of-Astronomical-Telescopes-Instruments-and-Systems/volume-8/issue-1/011021/Highlights-of-the-SKA1-Mid-telescope-architecture/10.1117/1.JATIS.8.1.011021.full},
    doi = {10.1117/1.JATIS.8.1.011021},
    number = {1},
    urldate = {2025-10-03},
    journal = {Journal of Astronomical Telescopes, Instruments, and Systems},
    author = {Swart, Gerhard P. and Dewdney, Peter E. and Cremonini, Andrea},
    month = jan,
    year = {2022},
    note = {Publisher: SPIE},
    pages = {011021},
}

@inproceedings{selina_next-generation_2018,
    address = {Austin, United States},
    title = {The {Next}-{Generation} {Very} {Large} {Array}: a technical overview},
    isbn = {978-1-5106-1953-1 978-1-5106-1954-8},
    shorttitle = {The {Next}-{Generation} {Very} {Large} {Array}},
    url = {https://www.spiedigitallibrary.org/conference-proceedings-of-spie/10700/2312089/The-Next-Generation-Very-Large-Array-a-technical-overview/10.1117/12.2312089.full},
    doi = {10.1117/12.2312089},
    language = {en},
    urldate = {2025-10-03},
    booktitle = {Ground-based and {Airborne} {Telescopes} {VII}},
    publisher = {SPIE},
    author = {Selina, Robert and McKinnon, Mark and Beasley, Anthony J. and Murphy, Eric and Carilli, Chris and Butler, Bryan and Clark, Barry and Erickson, Alan and Grammer, Wes and Jackson, James and Kent, Brian and Mason, Brian and Morgan, Matthew and Ojeda, Omar and Shillue, William and Sturgis, Silver and Urbain, Denis},
    editor = {Gilmozzi, Roberto and Marshall, Heather K. and Spyromilio, Jason},
    month = jul,
    year = {2018},
    pages = {55},
}

@incollection{schilizzi_evolution_2024,
    address = {Cham},
    title = {Evolution of the {SKA} {Science} {Case}},
    isbn = {978-3-031-51374-9},
    url = {https://doi.org/10.1007/978-3-031-51374-9\_5},
    language = {en},
    urldate = {2025-10-03},
    booktitle = {The {Square} {Kilometre} {Array}: {A} {Science} {Mega}-{Project} in the {Making}, 1990-2012},
    publisher = {Springer International Publishing},
    author = {Schilizzi, Richard T. and Ekers, Ronald D. and Dewdney, Peter E. and Crosby, Philip},
    editor = {Schilizzi, Richard T. and Ekers, Ronald D. and Dewdney, Peter E. and Crosby, Philip},
    year = {2024},
    doi = {10.1007/978-3-031-51374-9_5},
    pages = {199--278},
}

@article{collaboration_first_2019,
    title = {First {M87} {Event} {Horizon} {Telescope} {Results}. {I}. {The} {Shadow} of the {Supermassive} {Black} {Hole}},
    volume = {875},
    issn = {2041-8205},
    url = {https://doi.org/10.3847/2041-8213/ab0ec7},
    doi = {10.3847/2041-8213/ab0ec7},
    language = {en},
    number = {1},
    urldate = {2025-10-03},
    journal = {The Astrophysical Journal Letters},
    author = {Collaboration, The Event Horizon Telescope and Akiyama, Kazunori and Alberdi, Antxon and Alef, Walter and Asada, Keiichi and Azulay, Rebecca and Baczko, Anne-Kathrin and Ball, David and Baloković, Mislav and Barrett, John and Bintley, Dan and Blackburn, Lindy and Boland, Wilfred and Bouman, Katherine L. and Bower, Geoffrey C. and Bremer, Michael and Brinkerink, Christiaan D. and Brissenden, Roger and Britzen, Silke and Broderick, Avery E. and Broguiere, Dominique and Bronzwaer, Thomas and Byun, Do-Young and Carlstrom, John E. and Chael, Andrew and Chan, Chi-kwan and Chatterjee, Shami and Chatterjee, Koushik and Chen, Ming-Tang and Chen, Yongjun and Cho, Ilje and Christian, Pierre and Conway, John E. and Cordes, James M. and Crew, Geoffrey B. and Cui, Yuzhu and Davelaar, Jordy and Laurentis, Mariafelicia De and Deane, Roger and Dempsey, Jessica and Desvignes, Gregory and Dexter, Jason and Doeleman, Sheperd S. and Eatough, Ralph P. and Falcke, Heino and Fish, Vincent L. and Fomalont, Ed and Fraga-Encinas, Raquel and Freeman, William T. and Friberg, Per and Fromm, Christian M. and Gómez, José L. and Galison, Peter and Gammie, Charles F. and García, Roberto and Gentaz, Olivier and Georgiev, Boris and Goddi, Ciriaco and Gold, Roman and Gu, Minfeng and Gurwell, Mark and Hada, Kazuhiro and Hecht, Michael H. and Hesper, Ronald and Ho, Luis C. and Ho, Paul and Honma, Mareki and Huang, Chih-Wei L. and Huang, Lei and Hughes, David H. and Ikeda, Shiro and Inoue, Makoto and Issaoun, Sara and James, David J. and Jannuzi, Buell T. and Janssen, Michael and Jeter, Britton and Jiang, Wu and Johnson, Michael D. and Jorstad, Svetlana and Jung, Taehyun and Karami, Mansour and Karuppusamy, Ramesh and Kawashima, Tomohisa and Keating, Garrett K. and Kettenis, Mark and Kim, Jae-Young and Kim, Junhan and Kim, Jongsoo and Kino, Motoki and Koay, Jun Yi and Koch, Patrick M. and Koyama, Shoko and Kramer, Michael and Kramer, Carsten and Krichbaum, Thomas P. and Kuo, Cheng-Yu and Lauer, Tod R. and Lee, Sang-Sung and Li, Yan-Rong and Li, Zhiyuan and Lindqvist, Michael and Liu, Kuo and Liuzzo, Elisabetta and Lo, Wen-Ping and Lobanov, Andrei P. and Loinard, Laurent and Lonsdale, Colin and Lu, Ru-Sen and MacDonald, Nicholas R. and Mao, Jirong and Markoff, Sera and Marrone, Daniel P. and Marscher, Alan P. and Martí-Vidal, Iván and Matsushita, Satoki and Matthews, Lynn D. and Medeiros, Lia and Menten, Karl M. and Mizuno, Yosuke and Mizuno, Izumi and Moran, James M. and Moriyama, Kotaro and Moscibrodzka, Monika and Müller, Cornelia and Nagai, Hiroshi and Nagar, Neil M. and Nakamura, Masanori and Narayan, Ramesh and Narayanan, Gopal and Natarajan, Iniyan and Neri, Roberto and Ni, Chunchong and Noutsos, Aristeidis and Okino, Hiroki and Olivares, Héctor and Ortiz-León, Gisela N. and Oyama, Tomoaki and Özel, Feryal and Palumbo, Daniel C. M. and Patel, Nimesh and Pen, Ue-Li and Pesce, Dominic W. and Piétu, Vincent and Plambeck, Richard and PopStefanija, Aleksandar and Porth, Oliver and Prather, Ben and Preciado-López, Jorge A. and Psaltis, Dimitrios and Pu, Hung-Yi and Ramakrishnan, Venkatessh and Rao, Ramprasad and Rawlings, Mark G. and Raymond, Alexander W. and Rezzolla, Luciano and Ripperda, Bart and Roelofs, Freek and Rogers, Alan and Ros, Eduardo and Rose, Mel and Roshanineshat, Arash and Rottmann, Helge and Roy, Alan L. and Ruszczyk, Chet and Ryan, Benjamin R. and Rygl, Kazi L. J. and Sánchez, Salvador and Sánchez-Arguelles, David and Sasada, Mahito and Savolainen, Tuomas and Schloerb, F. Peter and Schuster, Karl-Friedrich and Shao, Lijing and Shen, Zhiqiang and Small, Des and Sohn, Bong Won and SooHoo, Jason and Tazaki, Fumie and Tiede, Paul and Tilanus, Remo P. J. and Titus, Michael and Toma, Kenji and Torne, Pablo and Trent, Tyler and Trippe, Sascha and Tsuda, Shuichiro and Bemmel, Ilse van and van Langevelde, Huib Jan and van Rossum, Daniel R. and Wagner, Jan and Wardle, John and Weintroub, Jonathan and Wex, Norbert and Wharton, Robert and Wielgus, Maciek and Wong, George N. and Wu, Qingwen and Young, Ken and Young, André and Younsi, Ziri and Yuan, Feng and Yuan, Ye-Fei and Zensus, J. Anton and Zhao, Guangyao and Zhao, Shan-Shan and Zhu, Ziyan and Algaba, Juan-Carlos and Allardi, Alexander and Amestica, Rodrigo and Anczarski, Jadyn and Bach, Uwe and Baganoff, Frederick K. and Beaudoin, Christopher and Benson, Bradford A. and Berthold, Ryan and Blanchard, Jay M. and Blundell, Ray and Bustamente, Sandra and Cappallo, Roger and Castillo-Domínguez, Edgar and Chang, Chih-Cheng and Chang, Shu-Hao and Chang, Song-Chu and Chen, Chung-Chen and Chilson, Ryan and Chuter, Tim C. and Rosado, Rodrigo Córdova and Coulson, Iain M. and Crawford, Thomas M. and Crowley, Joseph and David, John and Derome, Mark and Dexter, Matthew and Dornbusch, Sven and Dudevoir, Kevin A. and Dzib, Sergio A. and Eckart, Andreas and Eckert, Chris and Erickson, Neal R. and Everett, Wendeline B. and Faber, Aaron and Farah, Joseph R. and Fath, Vernon and Folkers, Thomas W. and Forbes, David C. and Freund, Robert and Gómez-Ruiz, Arturo I. and Gale, David M. and Gao, Feng and Geertsema, Gertie and Graham, David A. and Greer, Christopher H. and Grosslein, Ronald and Gueth, Frédéric and Haggard, Daryl and Halverson, Nils W. and Han, Chih-Chiang and Han, Kuo-Chang and Hao, Jinchi and Hasegawa, Yutaka and Henning, Jason W. and Hernández-Gómez, Antonio and Herrero-Illana, Rubén and Heyminck, Stefan and Hirota, Akihiko and Hoge, James and Huang, Yau-De and Impellizzeri, C. M. Violette and Jiang, Homin and Kamble, Atish and Keisler, Ryan and Kimura, Kimihiro and Kono, Yusuke and Kubo, Derek and Kuroda, John and Lacasse, Richard and Laing, Robert A. and Leitch, Erik M. and Li, Chao-Te and Lin, Lupin C.-C. and Liu, Ching-Tang and Liu, Kuan-Yu and Lu, Li-Ming and Marson, Ralph G. and Martin-Cocher, Pierre L. and Massingill, Kyle D. and Matulonis, Callie and McColl, Martin P. and McWhirter, Stephen R. and Messias, Hugo and Meyer-Zhao, Zheng and Michalik, Daniel and Montaña, Alfredo and Montgomerie, William and Mora-Klein, Matias and Muders, Dirk and Nadolski, Andrew and Navarro, Santiago and Neilsen, Joseph and Nguyen, Chi H. and Nishioka, Hiroaki and Norton, Timothy and Nowak, Michael A. and Nystrom, George and Ogawa, Hideo and Oshiro, Peter and Oyama, Tomoaki and Parsons, Harriet and Paine, Scott N. and Peñalver, Juan and Phillips, Neil M. and Poirier, Michael and Pradel, Nicolas and Primiani, Rurik A. and Raffin, Philippe A. and Rahlin, Alexandra S. and Reiland, George and Risacher, Christopher and Ruiz, Ignacio and Sáez-Madaín, Alejandro F. and Sassella, Remi and Schellart, Pim and Shaw, Paul and Silva, Kevin M. and Shiokawa, Hotaka and Smith, David R. and Snow, William and Souccar, Kamal and Sousa, Don and Sridharan, T. K. and Srinivasan, Ranjani and Stahm, William and Stark, Anthony A. and Story, Kyle and Timmer, Sjoerd T. and Vertatschitsch, Laura and Walther, Craig and Wei, Ta-Shun and Whitehorn, Nathan and Whitney, Alan R. and Woody, David P. and Wouterloot, Jan G. A. and Wright, Melvin and Yamaguchi, Paul and Yu, Chen-Yu and Zeballos, Milagros and Zhang, Shuo and Ziurys, Lucy},
    month = apr,
    year = {2019},
    note = {Publisher: The American Astronomical Society},
    pages = {L1},
}

@article{sovers_astrometry_1998,
    title = {Astrometry and geodesy with radio interferometry: experiments, models, results},
    volume = {70},
    shorttitle = {Astrometry and geodesy with radio interferometry},
    url = {https://link.aps.org/doi/10.1103/RevModPhys.70.1393},
    doi = {10.1103/RevModPhys.70.1393},
    number = {4},
    urldate = {2025-10-03},
    journal = {Reviews of Modern Physics},
    author = {Sovers, Ojars J. and Fanselow, John L. and Jacobs, Christopher S.},
    month = oct,
    year = {1998},
    note = {Publisher: American Physical Society},
    pages = {1393--1454},
}

@article{lonsdale_murchison_2009,
    title = {The {Murchison} {Widefield} {Array}: {Design} {Overview}},
    volume = {97},
    copyright = {https://ieeexplore.ieee.org/Xplorehelp/downloads/license-information/IEEE.html},
    issn = {0018-9219, 1558-2256},
    shorttitle = {The {Murchison} {Widefield} {Array}},
    url = {http://ieeexplore.ieee.org/document/5164979/},
    doi = {10.1109/JPROC.2009.2017564},
    language = {en},
    number = {8},
    urldate = {2025-09-24},
    journal = {Proceedings of the IEEE},
    author = {Lonsdale, C.J. and Cappallo, R.J. and Morales, M.F. and Briggs, F.H. and Benkevitch, L. and Bowman, J.D. and Bunton, J.D. and Burns, S. and Corey, B.E. and deSouza, L. and Doeleman, S.S. and Derome, M. and Deshpande, A. and Gopala, M.R. and Greenhill, L.J. and Herne, D.E. and Hewitt, J.N. and Kamini, P.A. and Kasper, J.C. and Kincaid, B.B. and Kocz, J. and Kowald, E. and Kratzenberg, E. and Kumar, D. and Lynch, M.J. and Madhavi, S. and Matejek, M. and Mitchell, D.A. and Morgan, E. and Oberoi, D. and Ord, S. and Pathikulangara, J. and Prabu, T. and Rogers, A. and Roshi, A. and Salah, J.E. and Sault, R.J. and Shankar, N.U. and Srivani, K.S. and Stevens, J. and Tingay, S. and Vaccarella, A. and Waterson, M. and Wayth, R.B. and Webster, R.L. and Whitney, A.R. and Williams, A. and Williams, C.},
    month = aug,
    year = {2009},
    pages = {1497--1506},
}

@book{schilizzi_square_2024,
    address = {Cham},
    series = {Historical \& {Cultural} {Astronomy}},
    title = {The {Square} {Kilometre} {Array}: {A} {Science} {Mega}-{Project} in the {Making}, 1990-2012},
    copyright = {https://creativecommons.org/licenses/by/4.0},
    isbn = {978-3-031-51373-2 978-3-031-51374-9},
    shorttitle = {The {Square} {Kilometre} {Array}},
    url = {https://link.springer.com/10.1007/978-3-031-51374-9},
    language = {en},
    urldate = {2025-10-03},
    publisher = {Springer International Publishing},
    author = {Schilizzi, Richard T. and Ekers, Ronald D. and Dewdney, Peter E. and Crosby, Philip},
    year = {2024},
    doi = {10.1007/978-3-031-51374-9},
}

@article{selina_ngvla_nodate,
	title = {The {ngVLA} {Reference} {Design}},
	language = {en},
	author = {Selina, Robert J and Murphy, Eric J and McKinnon, Mark and Beasley, Anthony and Butler, Bryan and Carilli, Chris and Clark, Barry and Durand, Steven and Erickson, Alan and Grammer, Wes and Hiriart, Rafael and Jackson, James and Kent, Brian and Mason, Brian and Morgan, Matthew and Ojeda, Omar Yeste and Rosero, Viviana and Shillue, William and Sturgis, Silver and Urbain, Denis},
}

@article{muir_road_2025,
    title = {The road to commercial success for neuromorphic technologies},
    volume = {16},
    copyright = {2025 The Author(s)},
    issn = {2041-1723},
    url = {https://www.nature.com/articles/s41467-025-57352-1},
    doi = {10.1038/s41467-025-57352-1},
    language = {en},
    number = {1},
    urldate = {2025-06-09},
    journal = {Nature Communications},
    author = {Muir, Dylan Richard and Sheik, Sadique},
    month = apr,
    year = {2025},
    note = {Publisher: Nature Publishing Group},
    pages = {3586},
}

@article{line_verifying_2024,
    title = {Verifying the {Australian} {MWA} {EoR} pipeline {I}: 21-cm sky model and correlated measurement density},
    volume = {41},
    issn = {1323-3580, 1448-6083},
    shorttitle = {Verifying the {Australian} {MWA} {EoR} pipeline {I}},
    url = {https://www.cambridge.org/core/journals/publications-of-the-astronomical-society-of-australia/article/verifying-the-australian-mwa-eor-pipeline-i-21cm-sky-model-and-correlated-measurement-density/61AEF74A93A547D2BF13C0CC8A8FFB59},
    doi = {10.1017/pasa.2024.31},
    language = {en},
    urldate = {2025-10-03},
    journal = {Publications of the Astronomical Society of Australia},
    author = {Line, J. L. B. and Trott, C. M. and Cook, J. H. and Greig, B. and Barry, N. and Jordan, C. H.},
    month = jan,
    year = {2024},
    pages = {e067},
}

@article{hurley-walker_radio_2022,
    title = {A radio transient with unusually slow periodic emission},
    volume = {601},
    copyright = {2022 The Author(s), under exclusive licence to Springer Nature Limited},
    issn = {1476-4687},
    url = {https://www.nature.com/articles/s41586-021-04272-x},
    doi = {10.1038/s41586-021-04272-x},
    language = {en},
    number = {7894},
    urldate = {2025-10-03},
    journal = {Nature},
    author = {Hurley-Walker, N. and Zhang, X. and Bahramian, A. and McSweeney, S. J. and O’Doherty, T. N. and Hancock, P. J. and Morgan, J. S. and Anderson, G. E. and Heald, G. H. and Galvin, T. J.},
    month = jan,
    year = {2022},
    note = {Publisher: Nature Publishing Group},
    pages = {526--530},
}

@article{wayth_gleam_2015,
    title = {{GLEAM}: {The} {GaLactic} and {Extragalactic} {All}-{Sky} {MWA} {Survey}},
    volume = {32},
    issn = {1323-3580, 1448-6083},
    shorttitle = {{GLEAM}},
    url = {https://www.cambridge.org/core/journals/publications-of-the-astronomical-society-of-australia/article/gleam-the-galactic-and-extragalactic-allsky-mwa-survey/31003E4628253D7A7BECBE5BCA07DC5D},
    doi = {10.1017/pasa.2015.26},
    language = {en},
    urldate = {2025-10-03},
    journal = {Publications of the Astronomical Society of Australia},
    author = {Wayth, R. B. and Lenc, E. and Bell, M. E. and Callingham, J. R. and Dwarakanath, K. S. and Franzen, T. M. O. and For, B.-Q. and Gaensler, B. and Hancock, P. and Hindson, L. and Hurley-Walker, N. and Jackson, C. A. and Johnston-Hollitt, M. and Kapińska, A. D. and McKinley, B. and Morgan, J. and Offringa, A. R. and Procopio, P. and Staveley-Smith, L. and Wu, C. and Zheng, Q. and Trott, C. M. and Bernardi, G. and Bowman, J. D. and Briggs, F. and Cappallo, R. J. and Corey, B. E. and Deshpande, A. A. and Emrich, D. and Goeke, R. and Greenhill, L. J. and Hazelton, B. J. and Kaplan, D. L. and Kasper, J. C. and Kratzenberg, E. and Lonsdale, C. J. and Lynch, M. J. and McWhirter, S. R. and Mitchell, D. A. and Morales, M. F. and Morgan, E. and Oberoi, D. and Ord, S. M. and Prabu, T. and Rogers, A. E. E. and Roshi, A. and Shankar, N. Udaya and Srivani, K. S. and Subrahmanyan, R. and Tingay, S. J. and Waterson, M. and Webster, R. L. and Whitney, A. R. and Williams, A. and Williams, C. L.},
    month = jan,
    year = {2015},
    pages = {e025},
}

@article{tingay_murchison_2013,
    title = {The {Murchison} {Widefield} {Array}: solar science with the low frequency {SKA} {Precursor}},
    volume = {440},
    copyright = {http://iopscience.iop.org/info/page/text-and-data-mining},
    issn = {1742-6596},
    shorttitle = {The {Murchison} {Widefield} {Array}},
    url = {https://iopscience.iop.org/article/10.1088/1742-6596/440/1/012033},
    doi = {10.1088/1742-6596/440/1/012033},
    language = {en},
    urldate = {2025-10-03},
    journal = {Journal of Physics: Conference Series},
    author = {Tingay, S J and Oberoi, D and Cairns, I and Donea, A and Duffin, R and Arcus, W and Bernardi, G and Bowman, J D and Briggs, F and Bunton, J D and Cappallo, R J and Corey, B E and Deshpande, A and deSouza, L and Emrich, D and Gaensler, B M and Goeke, R and Greenhill, L J and Hazelton, B J and Herne, D and Hewitt, J N and Johnston-Hollitt, M and Kaplan, D L and Kasper, J C and Kennewell, J A and Kincaid, B B and Koenig, R and Kratzenberg, E and Lonsdale, C J and Lynch, M J and McWhirter, S R and Mitchell, D A and Morales, M F and Morgan, E and Ord, S M and Pathikulangara, J and Prabu, T and Remillard, R A and Rogers, A E E and Roshi, A and Salah, J E and Sault, R J and Udaya-Shankar, N and Srivani, K S and Stevens, J and Subrahmanyan, R and Waterson, M and Wayth, R B and Webster, R L and Whitney, A R and Williams, A and Williams, C L and Wyithe, J S B},
    month = jun,
    year = {2013},
    pages = {012033},
}

@misc{noauthor_setonix_nodate,
    title = {Setonix},
    url = {https://pawsey.org.au/systems/setonix/},
    language = {en-US},
    urldate = {2025-10-03},
    journal = {Pawsey Supercomputing Research Centre},
}

@misc{noauthor_mwatelescopebirli_2025,
    title = {{MWATelescope}/{Birli}},
    copyright = {MPL-2.0},
    url = {https://github.com/MWATelescope/Birli},
    urldate = {2025-10-03},
    publisher = {Murchison Widefield Array (MWA) Telescope},
    month = sep,
    year = {2025},
    note = {original-date: 2021-01-20T05:53:56Z},
}

@article{duchesne_rapid_2025,
    title = {The {Rapid} {ASKAP} {Continuum} {Survey} ({RACS}) {VI}: {The} {RACS}-high 1 655.5 {MHz} images and catalogue},
    volume = {42},
    issn = {1323-3580, 1448-6083},
    shorttitle = {The {Rapid} {ASKAP} {Continuum} {Survey} ({RACS}) {VI}},
    url = {https://www.cambridge.org/core/journals/publications-of-the-astronomical-society-of-australia/article/rapid-askap-continuum-survey-racs-vi-the-racshigh-1-6555-mhz-images-and-catalogue/6961A97FA79D7E87143DF196F8EA53E3},
    doi = {10.1017/pasa.2025.2},
    language = {en},
    urldate = {2025-10-03},
    journal = {Publications of the Astronomical Society of Australia},
    author = {Duchesne, S. W. and Ross, K. and Thomson, A. J. M. and Lenc, E. and Murphy, Tara and Galvin, T. J. and Hotan, A. W. and Moss, V. and Whiting, Matthew T.},
    month = jan,
    year = {2025},
    pages = {e038},
}

@article{pingel_gaskap-hi_2022,
    title = {{GASKAP}-{HI} pilot survey science {I}: {ASKAP} zoom observations of {Hi} emission in the {Small} {Magellanic} {Cloud}},
    volume = {39},
    issn = {1323-3580, 1448-6083},
    shorttitle = {{GASKAP}-{HI} pilot survey science {I}},
    url = {https://www.cambridge.org/core/journals/publications-of-the-astronomical-society-of-australia/article/gaskaphi-pilot-survey-science-i-askap-zoom-observations-of-hi-emission-in-the-small-magellanic-cloud/3BFFB8CB77C0DB54F8E620E73E5B9805},
    doi = {10.1017/pasa.2021.59},
    language = {en},
    urldate = {2025-10-03},
    journal = {Publications of the Astronomical Society of Australia},
    author = {Pingel, N. M. and Dempsey, J. and McClure-Griffiths, N. M. and Dickey, J. M. and Jameson, K. E. and Arce, H. and Anglada, G. and Bland-Hawthorn, J. and Breen, S. L. and Buckland-Willis, F. and Clark, S. E. and Dawson, J. R. and Dénes, H. and Teodoro, E. M. Di and For, B.-Q. and Foster, Tyler J. and Gómez, J. F. and Imai, H. and Joncas, G. and Kim, C.-G. and Lee, M.-Y. and Lynn, C. and Leahy, D. and Ma, Y. K. and Marchal, A. and McConnell, D. and Miville-Deschènes, M.-A. and Moss, V. A. and Murray, C. E. and Nidever, D. and Peek, J. and Stanimirović, S. and Staveley-Smith, L. and Tepper-Garcia, T. and Tremblay, C. D. and Uscanga, L. and Loon, J. Th van and Vázquez-Semadeni, E. and Allison, J. R. and Anderson, C. S. and Ball, Lewis and Bell, M. and Bock, D. C.-J. and Bunton, J. and Cooray, F. R. and Cornwell, T. and Koribalski, B. S. and Gupta, N. and Hayman, D. B. and Harvey-Smith, L. and Lee-Waddell, K. and Ng, A. and Phillips, C. J. and Voronkov, M. and Westmeier, T. and Whiting, M. T.},
    month = jan,
    year = {2022},
    pages = {e005},
}

@article{hobbs_pilot_2016,
    title = {A pilot {ASKAP} survey of radio transient events in the region around the intermittent pulsar {PSR} {J1107}-5907},
    volume = {456},
    issn = {0035-8711},
    url = {https://doi.org/10.1093/mnras/stv2893},
    doi = {10.1093/mnras/stv2893},
    number = {4},
    urldate = {2025-10-03},
    journal = {Monthly Notices of the Royal Astronomical Society},
    author = {Hobbs, G. and Heywood, I. and Bell, M. E. and Kerr, M. and Rowlinson, A. and Johnston, S. and Shannon, R. M. and Voronkov, M. A. and Ward, C. and Banyer, J. and Hancock, P. J. and Murphy, Tara and Allison, J. R. and Amy, S. W. and Ball, L. and Bannister, K. and Bock, D. C.-J. and Brodrick, D. and Brothers, M. and Brown, A. J. and Bunton, J. D. and Chapman, J. and Chippendale, A. P. and Chung, Y. and DeBoer, D. and Diamond, P. and Edwards, P. G. and Ekers, R. and Ferris, R. H. and Forsyth, R. and Gough, R. and Grancea, A. and Gupta, N. and Harvey-Smith, L. and Hay, S. and Hayman, D. B. and Hotan, A. W. and Hoyle, S. and Humphreys, B. and Indermuehle, B. and Jacka, C. E. and Jackson, C. A. and Jackson, S. and Jeganathan, K. and Joseph, J and Kendall, R. and Kiraly, D. and Koribalski, B. and Leach, M. and Lenc, E. and MacLeod, A. and Mader, S. and Marquarding, M. and Marvil, J. and McClure-Griffiths, N. and McConnell, D. and Mirtschin, P. and Neuhold, S. and Ng, A. and Norris, R. P. and O'Sullivan, J. and Pearce, S. and Phillips, C. J. and Popping, A. and Qiao, R. Y. and Reynolds, J. E. and Roberts, P. and Sault, R. J. and Schinckel, A. E. T. and Serra, P. and Shaw, R. and Shimwell, T. W. and Storey, M. and Sweetnam, A. W. and Tzioumis, A. and Westmeier, T. and Whiting, M. and Wilson, C. D.},
    month = mar,
    year = {2016},
    pages = {3948--3960},
}

@misc{guzman_askap_2019,
    title = {{ASKAP} {Science} {Data} {Processor} software - {ASKAPsoft} {Version} 0.23.3},
    url = {https://data.csiro.au/collections/#collection/CIcsiro:39526v1/DItrue},
    urldate = {2025-10-03},
    publisher = {CSIRO},
    author = {Guzman, Juan and Whiting, Matthew and Voronkov, Max and Mitchell, Daniel and Ord, Stephen and Collins, Daniel and Marquarding, Malte and Lahur, Paulus and Maher, Tony and Van Diepen, Ger and Bannister, Keith and Wu, Xinyu and Lenc, Emil and Khoo, Jonathan and Bastholm, Eric},
    year = {2019},
    doi = {10.25919/5CCA3787A6353},
}

@article{offringa_wsclean_2014,
    title = {wsclean: an implementation of a fast, generic wide-field imager for radio astronomy},
    volume = {444},
    issn = {0035-8711},
    shorttitle = {wsclean},
    url = {https://doi.org/10.1093/mnras/stu1368},
    doi = {10.1093/mnras/stu1368},
    number = {1},
    urldate = {2025-10-03},
    journal = {Monthly Notices of the Royal Astronomical Society},
    author = {Offringa, A. R. and McKinley, B. and Hurley-Walker, N. and Briggs, F. H. and Wayth, R. B. and Kaplan, D. L. and Bell, M. E. and Feng, L. and Neben, A. R. and Hughes, J. D. and Rhee, J. and Murphy, T. and Bhat, N. D. R. and Bernardi, G. and Bowman, J. D. and Cappallo, R. J. and Corey, B. E. and Deshpande, A. A. and Emrich, D. and Ewall-Wice, A. and Gaensler, B. M. and Goeke, R. and Greenhill, L. J. and Hazelton, B. J. and Hindson, L. and Johnston-Hollitt, M. and Jacobs, D. C. and Kasper, J. C. and Kratzenberg, E. and Lenc, E. and Lonsdale, C. J. and Lynch, M. J. and McWhirter, S. R. and Mitchell, D. A. and Morales, M. F. and Morgan, E. and Kudryavtseva, N. and Oberoi, D. and Ord, S. M. and Pindor, B. and Procopio, P. and Prabu, T. and Riding, J. and Roshi, D. A. and Shankar, N. Udaya and Srivani, K. S. and Subrahmanyan, R. and Tingay, S. J. and Waterson, M. and Webster, R. L. and Whitney, A. R. and Williams, A. and Williams, C. L.},
    month = oct,
    year = {2014},
    pages = {606--619},
}

@incollection{braun_advancing_2015,
    title = {Advancing {Astrophysics} with the {Square} {Kilometre} {Array}},
    volume = {215},
    url = {https://pos.sissa.it/215/174/},
    language = {en},
    urldate = {2025-10-03},
    booktitle = {Proceedings of {Advancing} {Astrophysics} with the {Square} {Kilometre} {Array} — {PoS}({AASKA14})},
    publisher = {SISSA Medialab},
    author = {Braun, Robert and Bourke, T. L. and Green, James A. and Keane, Evan and Wagg, Jeff},
    month = may,
    year = {2015},
    doi = {10.22323/1.215.0174},
    note = {Conference Name: Advancing Astrophysics with the Square Kilometre Array},
    pages = {174},
}

@inproceedings{wang_processing_2020,
    title = {Processing {Full}-{Scale} {Square} {Kilometre} {Array} {Data} on the {Summit} {Supercomputer}},
    url = {https://ieeexplore.ieee.org/abstract/document/9355269},
    doi = {10.1109/SC41405.2020.00006},
    urldate = {2025-10-03},
    booktitle = {{SC20}: {International} {Conference} for {High} {Performance} {Computing}, {Networking}, {Storage} and {Analysis}},
    author = {Wang, Ruonan and Tobar, Rodrigo and Dolensky, Markus and An, Tao and Wicenec, Andreas and Wu, Chen and Dulwich, Fred and Podhorszki, Norbert and Anantharaj, Valentine and Suchyta, Eric and Lao, Baoqiang and Klasky, Scott},
    month = nov,
    year = {2020},
    pages = {1--12},
}

@techreport{noauthor_xylo-audio_2022,
    type = {Technical {Datasheet}},
    title = {Xylo-{Audio} {Development} {Kit} {Datasheet}},
    shorttitle = {Xylo-{Audio} {Datasheet}},
    url = {https://www.synsense.ai/wp-content/uploads/2023/06/Xylo-Audio-datasheet.pdf},
    language = {English},
    institution = {SynSense AG},
    month = jun,
    year = {2022},
}

@misc{gonzalez_spinnaker2_2024,
    title = {{SpiNNaker2}: {A} {Large}-{Scale} {Neuromorphic} {System} for {Event}-{Based} and {Asynchronous} {Machine} {Learning}},
    shorttitle = {{SpiNNaker2}},
    url = {http://arxiv.org/abs/2401.04491},
    doi = {10.48550/arXiv.2401.04491},
    language = {en},
    urldate = {2025-10-02},
    publisher = {arXiv},
    author = {Gonzalez, Hector A. and Huang, Jiaxin and Kelber, Florian and Nazeer, Khaleelulla Khan and Langer, Tim and Liu, Chen and Lohrmann, Matthias and Rostami, Amirhossein and Schöne, Mark and Vogginger, Bernhard and Wunderlich, Timo C. and Yan, Yexin and Akl, Mahmoud and Mayr, Christian},
    month = jan,
    year = {2024},
    note = {arXiv:2401.04491 [cs]},
}

@inproceedings{orchard_efficient_2021,
    title = {Efficient {Neuromorphic} {Signal} {Processing} with {Loihi} 2},
    url = {https://ieeexplore.ieee.org/abstract/document/9605018},
    doi = {10.1109/SiPS52927.2021.00053},
    urldate = {2024-10-22},
    booktitle = {2021 {IEEE} {Workshop} on {Signal} {Processing} {Systems} ({SiPS})},
    author = {Orchard, Garrick and Frady, E. Paxon and Rubin, Daniel Ben Dayan and Sanborn, Sophia and Shrestha, Sumit Bam and Sommer, Friedrich T. and Davies, Mike},
    month = oct,
    year = {2021},
    note = {ISSN: 2374-7390},
    pages = {254--259},
}

@misc{abreu_neuromorphic_2025,
    title = {Neuromorphic {Principles} for {Efficient} {Large} {Language} {Models} on {Intel} {Loihi} 2},
    url = {http://arxiv.org/abs/2503.18002},
    doi = {10.48550/arXiv.2503.18002},
    urldate = {2025-06-13},
    publisher = {arXiv},
    author = {Abreu, Steven and Shrestha, Sumit Bam and Zhu, Rui-Jie and Eshraghian, Jason},
    month = mar,
    year = {2025},
    note = {arXiv:2503.18002 [cs]},
}

@misc{brehove_sigma-delta_2025,
    title = {Sigma-{Delta} {Neural} {Network} {Conversion} on {Loihi} 2},
    url = {http://arxiv.org/abs/2505.06417},
    doi = {10.48550/arXiv.2505.06417},
    language = {en},
    urldate = {2025-10-02},
    publisher = {arXiv},
    author = {Brehove, Matthew and Tumpa, Sadia Anjum and Kyubwa, Espoir and Menon, Naresh and Narayanan, Vijaykrishnan},
    month = may,
    year = {2025},
    note = {arXiv:2505.06417 [cs]},
}

@inproceedings{shrestha_efficient_2024,
    title = {Efficient {Video} and {Audio} {Processing} with {Loihi} 2},
    url = {https://ieeexplore.ieee.org/abstract/document/10448003},
    doi = {10.1109/ICASSP48485.2024.10448003},
    urldate = {2025-09-25},
    booktitle = {{ICASSP} 2024 - 2024 {IEEE} {International} {Conference} on {Acoustics}, {Speech} and {Signal} {Processing} ({ICASSP})},
    author = {Shrestha, Sumit Bam and Timcheck, Jonathan and Frady, Paxon and Campos-Macias, Leobardo and Davies, Mike},
    month = apr,
    year = {2024},
    note = {ISSN: 2379-190X},
    pages = {13481--13485},
}

@misc{corda_near_2020,
    title = {Near {Memory} {Acceleration} on {High} {Resolution} {Radio} {Astronomy} {Imaging}},
    url = {http://arxiv.org/abs/2005.04098},
    doi = {10.48550/arXiv.2005.04098},
    language = {en},
    urldate = {2025-10-03},
    publisher = {arXiv},
    author = {Corda, Stefano and Veenboer, Bram and Awan, Ahsan Javed and Kumar, Akash and Jordans, Roel and Corporaal, Henk},
    month = may,
    year = {2020},
    note = {arXiv:2005.04098 [cs]},
}

@article{fiorin_exploring_2016,
    title = {Exploring the {Design} {Space} of an {Energy}-{Efficient} {Accelerator} for the {SKA1}-{Low} {Central} {Signal} {Processor}},
    volume = {44},
    issn = {1573-7640},
    url = {https://doi.org/10.1007/s10766-016-0420-y},
    doi = {10.1007/s10766-016-0420-y},
    language = {en},
    number = {5},
    urldate = {2025-10-06},
    journal = {International Journal of Parallel Programming},
    author = {Fiorin, Leandro and Vermij, Erik and van Lunteren, Jan and Jongerius, Rik and Hagleitner, Christoph},
    month = oct,
    year = {2016},
    pages = {1003--1027},
}

@inproceedings{jonas_meerkat_2018,
    address = {Stellenbosch, South Africa},
    title = {The {MeerKAT} {Radio} {Telescope}},
    url = {https://pos.sissa.it/277/001},
    doi = {10.22323/1.277.0001},
    language = {en},
    urldate = {2025-10-06},
    booktitle = {Proceedings of {MeerKAT} {Science}: {On} the {Pathway} to the {SKA} — {PoS}({MeerKAT2016})},
    publisher = {Sissa Medialab},
    author = {Jonas, Justin and {the MeerKAT Team}},
    month = feb,
    year = {2018},
    pages = {001},
}

@inproceedings{raicu_many-task_2008,
    title = {Many-task computing for grids and supercomputers},
    url = {https://ieeexplore.ieee.org/document/4777912},
    doi = {10.1109/MTAGS.2008.4777912},
    urldate = {2025-10-06},
    booktitle = {2008 {Workshop} on {Many}-{Task} {Computing} on {Grids} and {Supercomputers}},
    author = {Raicu, Ioan and Foster, Ian T. and Zhao, Yong},
    month = nov,
    year = {2008},
    note = {ISSN: 2151-1691},
    pages = {1--11},
}

@article{wu_daliuge:_2017,
	title = {{DALiuGE}: {A} graph execution framework for harnessing the astronomical data deluge},
	volume = {20},
	issn = {22131337},
	shorttitle = {{DALiuGE}},
	url = {https://linkinghub.elsevier.com/retrieve/pii/S2213133716301214},
	doi = {10.1016/j.ascom.2017.03.007},
	language = {en},
	urldate = {2019-07-17},
	journal = {Astronomy and Computing},
	author = {Wu, C. and Tobar, R. and Vinsen, K. and Wicenec, A. and Pallot, D. and Lao, B. and Wang, R. and An, T. and Boulton, M. and Cooper, I. and Dodson, R. and Dolensky, M. and Mei, Y. and Wang, F.},
	month = jul,
	year = {2017},
	pages = {1--15},
}

@techreport{advanced_micro_devices_amd_2020,
    title = {{AMD} {Epyc} 7002 {Series} {Processors}},
    url = {https://www.amd.com/content/dam/amd/en/documents/products/epyc/amd-epyc-7002-series-datasheet.pdf},
    author = {Advanced Micro Devices},
    month = apr,
    year = {2020},
}

@techreport{advanced_micro_devices_alveo_2023,
    title = {Alveo {U280} {Data} {Center} {Accelerator} {Card} {Data} {Sheet} ({DS963})},
    url = {https://docs.amd.com/r/en-US/ds963-u280},
    author = {Advanced Micro Devices},
    month = jun,
    year = {2023},
}

@techreport{nvidia_corporation_nvidia_2020,
    title = {{NVIDIA} {V100} {Tensor} {Core} {GPU}},
    url = {https://images.nvidia.com/content/technologies/volta/pdf/volta-v100-datasheet-update-us-1165301-r5.pdf},
    author = {NVidia Corporation},
    month = jan,
    year = {2020},
}

@techreport{intel_intel_2017,
    title = {Intel {Xeon} {Gold} 6140 {Processor} {Datasheet}},
    url = {https://www.intel.com/content/www/us/en/products/sku/120485/intel-xeon-gold-6140-processor-24-75m-cache-2-30-ghz/specifications.html},
    author = {Intel},
    month = jul,
    year = {2017},
}

@techreport{intel_intel_2014,
    title = {Intel {Xeon} {Processor} {E5}-2680 v3 {Datasheet}},
    url = {https://www.intel.com/content/www/us/en/products/sku/81908/intel-xeon-processor-e52680-v3-30m-cache-2-50-ghz/specifications.html},
    author = {Intel},
    month = sep,
    year = {2014},
}

@techreport{intel_intel_2014_2,
	title = {Intel {Xeon} {Processor} {E5}-2630 v3 {Datasheet}},
	url = {https://www.intel.com/content/www/us/en/products/sku/83356/intel-xeon-processor-e52630-v3-20m-cache-2-40-ghz/specifications.html},
	author = {Intel},
	month = sep,
	year = {2014},
}

@techreport{nvidia_corporation_nvidia_2013,
	title = {{NVIDIA} {Tesla} {K20X} {GPU} {Accelerator}},
	url = {https://www.nvidia.com/content/PDF/kepler/Tesla-K20X-BD-06397-001-v07.pdf},
	author = {NVidia Corporation},
	month = jul,
	year = {2013},
}

@techreport{spinncloud_spinnaker2_2025,
	title = {{SpiNNaker2} {Chip} {Topology}},
	url = {https://spinnaker2.gitlab.io/external/documentation/hardware/2-s2-chip-topology/#chip-overview},
	author = {SpiNNcloud},
	year = {2025},
}

@article{selina_ngvla_2024,
    title = {{ngVLA}: {Project} {Technical} {Overview}},
    language = {en},
    author = {Selina, Robert},
    year = {2024},
}

@inproceedings{pham_review_2021,
	address = {Hanoi, Vietnam},
	title = {A review of {SNN} implementation on {FPGA}},
	copyright = {https://ieeexplore.ieee.org/Xplorehelp/downloads/license-information/IEEE.html},
	isbn = {978-1-6654-1910-9},
	url = {https://ieeexplore.ieee.org/document/9585245/},
	doi = {10.1109/MAPR53640.2021.9585245},
	urldate = {2026-01-07},
	booktitle = {2021 {International} {Conference} on {Multimedia} {Analysis} and {Pattern} {Recognition} ({MAPR})},
	publisher = {IEEE},
	author = {Pham, Quoc Trung and Nguyen, Thu Quyen and Hoang, Phuong Chi and Dang, Quang Hieu and Nguyen, Duc Minh and Nguyen, Huy Hoang},
	month = oct,
	year = {2021},
	pages = {1--6},
}

@inproceedings{miniskar_neuro-spark_2024,
	address = {Arlington, VA, USA},
	title = {Neuro-{Spark}: {A} {Submicrosecond} {Spiking} {Neural} {Networks} {Architecture} for {In}-{Sensor} {Filtering}},
	copyright = {https://doi.org/10.15223/policy-029},
	isbn = {979-8-3503-6865-9},
	shorttitle = {Neuro-{Spark}},
	url = {https://ieeexplore.ieee.org/document/10766567/},
	doi = {10.1109/ICONS62911.2024.00017},
	urldate = {2026-01-07},
	booktitle = {2024 {International} {Conference} on {Neuromorphic} {Systems} ({ICONS})},
	publisher = {IEEE},
	author = {Miniskar, Narsinga Rao and Young, Aaron R. and Asifuzzaman, Kazi and Kulkarni, Shruti and Date, Prasanna and Bean, Alice and Vetter, Jeffrey S.},
	month = jul,
	year = {2024},
	pages = {63--70},
}

@article{heidarpur_cordic-snn_2019,
	title = {{CORDIC}-{SNN}: {On}-{FPGA} {STDP} {Learning} {With} {Izhikevich} {Neurons}},
	volume = {66},
	copyright = {https://ieeexplore.ieee.org/Xplorehelp/downloads/license-information/IEEE.html},
	issn = {1549-8328, 1558-0806},
	shorttitle = {{CORDIC}-{SNN}},
	url = {https://ieeexplore.ieee.org/document/8660497/},
	doi = {10.1109/TCSI.2019.2899356},
	number = {7},
	urldate = {2026-01-07},
	journal = {IEEE Transactions on Circuits and Systems I: Regular Papers},
	author = {Heidarpur, Moslem and Ahmadi, Arash and Ahmadi, Majid and Rahimi Azghadi, Mostafa},
	month = jul,
	year = {2019},
	pages = {2651--2661},
}

@article{young_review_2019,
	title = {A {Review} of {Spiking} {Neuromorphic} {Hardware} {Communication} {Systems}},
	volume = {7},
	copyright = {https://creativecommons.org/licenses/by/4.0/legalcode},
	issn = {2169-3536},
	url = {https://ieeexplore.ieee.org/document/8843969/},
	doi = {10.1109/ACCESS.2019.2941772},
	urldate = {2026-01-07},
	journal = {IEEE Access},
	author = {Young, Aaron R. and Dean, Mark E. and Plank, James S. and Rose, Garrett S.},
	year = {2019},
	pages = {135606--135620},
}

@article{eappen_neuromorphic_2025,
	title = {Neuromorphic {Models} for {Energy}-{Efficient} {Onboard} {Interference} {Detection} in {Satellite} {Systems}},
	volume = {61},
	copyright = {https://ieeexplore.ieee.org/Xplorehelp/downloads/license-information/IEEE.html},
	issn = {0018-9251, 1557-9603, 2371-9877},
	url = {https://ieeexplore.ieee.org/document/11155214/},
	doi = {10.1109/TAES.2025.3608092},
	number = {6},
	urldate = {2026-01-07},
	journal = {IEEE Transactions on Aerospace and Electronic Systems},
	author = {Eappen, Geoffrey and Daoud, Saed and Skatchkovsky, Nicolas and Ortiz, Flor and Lagunas, Eva and Martins, Wallace A. and Chatzinotas, Symeon},
	month = dec,
	year = {2025},
	pages = {17682--17702},
}

@mastersthesis{bos_analogue_2025,
    address = {Netherlands},
    title = {Analogue neuromorphic receiver signal processing for radio astronomy},
    language = {en},
    school = {Radboud University},
    author = {Bos, Hidde},
    month = jun,
    year = {2025},
}

@misc{eshraghian_training_2022,
    title = {Training {Spiking} {Neural} {Networks} {Using} {Lessons} {From} {Deep} {Learning}},
    url = {http://arxiv.org/abs/2109.12894},
    doi = {10.48550/arXiv.2109.12894},
    urldate = {2023-01-18},
    publisher = {arXiv},
    author = {Eshraghian, Jason K. and Ward, Max and Neftci, Emre and Wang, Xinxin and Lenz, Gregor and Dwivedi, Girish and Bennamoun, Mohammed and Jeong, Doo Seok and Lu, Wei D.},
    month = jan,
    year = {2022},
    note = {arXiv:2109.12894 [cs]},
}

@article{theilman_solving_2025,
    title = {Solving sparse finite element problems on neuromorphic hardware},
    volume = {7},
    issn = {2522-5839},
    url = {https://www.nature.com/articles/s42256-025-01143-2},
    doi = {10.1038/s42256-025-01143-2},
    language = {en},
    number = {11},
    urldate = {2026-01-09},
    journal = {Nature Machine Intelligence},
    author = {Theilman, Bradley H. and Aimone, James B.},
    month = nov,
    year = {2025},
    pages = {1845--1857},
}

\end{document}